\documentclass[iop,apj]{emulateapj}

\usepackage{amssymb}
\usepackage{amsmath}
\usepackage{graphicx}
\usepackage{rotating}
\usepackage{natbib}

\newcommand\peryr{${\rm yr}^{-1}$}

\shorttitle{OGLE Atlas of Classical Novae}
\shortauthors{Mr\'oz et al.}

\begin{document}

\title{OGLE Atlas of Classical Novae \\I. Galactic Bulge Objects}

\author{P. Mr{\'o}z\altaffilmark{1}, A. Udalski\altaffilmark{1}, R. Poleski\altaffilmark{1,2}, I. Soszy{\'n}ski\altaffilmark{1}, M.~K. Szyma{\'n}ski\altaffilmark{1}, G. Pietrzy{\'n}ski\altaffilmark{1,3}, \\ {\L{}}. Wyrzykowski\altaffilmark{1,4}, K. Ulaczyk\altaffilmark{1}, S. Koz{\l{}}owski\altaffilmark{1}, P. Pietrukowicz\altaffilmark{1}, and J. Skowron\altaffilmark{1}}
\email{pmroz@astrouw.edu.pl}
\altaffiltext{1}{Warsaw University Observatory, Al. Ujazdowskie 4, 00-478 Warszawa, Poland}
\altaffiltext{2}{Department of Astronomy, Ohio State University, 140 W. 18th Ave., Columbus, OH 43210, USA}
\altaffiltext{3}{Universidad de Concepci\'on, Departamento de Astronom\'ia, Casilla 160-C, Concepci\'on, Chile}
\altaffiltext{4}{Institute of Astronomy, University of Cambridge, Madingley Road, Cambridge CB3 0HA, UK}

\begin{abstract}
Eruptions of classical novae are possible sources of lithium formation and gamma-ray emission. Nova remnants can also become Type Ia supernovae (SNe Ia). The contribution of novae to these phenomena depends on nova rates, which are not well established for the Galaxy. Here, we directly measure a Galactic bulge nova rate of $13.8 \pm 2.6$ \peryr. This measurement is much more accurate than any previous measurement of this kind thanks to many years' monitoring of the bulge by the Optical Gravitational Lensing Experiment (OGLE) survey. Our sample consists of 39 novae eruptions, $\sim$1/3 of which are OGLE-based discoveries. This long-term monitoring allows us to not only measure the nova rate but also to study in detail the light curves of 39 eruptions and more than 80 post-nova candidates. We measured orbital periods for 9 post-novae and 9 novae, and in 14 cases we procured the first estimates. The OGLE survey is very sensitive to the frequently erupting recurrent novae. We did not find any object similar to M31 2008-12a, which erupts once a year. The lack of detection indicates that there is only a small number of them in the Galactic bulge.
\end{abstract}


\section{Introduction}

Eruptions of classical novae (CNe) occur in binary systems consisting of a white dwarf (primary) and a non-degenerate companion, usually a main-sequence star, subgiant, or red giant (secondary). The matter from the secondary is transferred -- via Roche lobe overflow or/and wind -- to the white dwarf surface. As the mass of accreted gas reaches a critical value, a thermonuclear runaway is triggered leading to the observed explosion. See Bode \& Evans (2008) for a review.

Nova eruptions are believed to recur every $10^3 - 10^6$ years. However, a small subgroup -- called recurrent novae (RNe) -- have shown at least two eruptions during the last $\sim 100$ years. They probably host extremely massive white dwarfs, and hence they are considered to be  potential progenitors of SNe Ia. In the inter-eruption time, novae behave as normal cataclysmic variables (CVs; Warner 2003). 

Studies of novae are important for a number of reasons. They are sources of high-energy gamma-rays (Ackermann et al. 2014; Chomiuk et al. 2014). They enrich the interstellar medium in heavy elements, including lithium (Tajitsu et al. 2015). Nova eruptions also provide a unique opportunity for studying the common-envelope process (Chomiuk et al. 2014).

Photometric observations from the Optical Gravitational Lensing Experiment (OGLE){\footnote{http://ogle.astrouw.edu.pl/} variability survey provide information about the overall population of CNe and enable in-depth studies of particularly interesting objects. This paper is divided into three main parts. In Section \ref{section:eruptions}, we present the results of a search for nova eruptions in the Galactic bulge. We describe in detail newly discovered systems, as well as those known from the literature. In Section \ref{sec:old}, we search for different types of variability in post-nova systems, occurring on timescales from minutes to decades: eclipses, secondary pulsations, fading events, dwarf nova outbursts, and RN eruptions.
Finally, in Sections \ref{sec:rate} and \ref{section:howmanyRN}, we calculate the Galactic bulge nova rate. We also provide estimates for a number of RNe with short inter-eruption timescales.

\section{Data}\label{sec:data}

All of the data presented in this paper were collected in the course of the OGLE. The OGLE sky survey began operation in 1992. In the first phase, OGLE-I, conducted from 1992 to 1995 with the 1.0 m Swope telescope at Las Campanas Observatory (LCO), Chile, 16 Galactic bulge fields with a total field of view of about 1 sq. degree were observed. The OGLE survey entered its second phase, OGLE-II, with the construction of a dedicated 1.3 m Warsaw Telescope (located at LCO). From 1997 to 2000, approximately 30 million stars from 11 sq. degrees of the Galactic bulge were monitored. The observing capabilities of the survey increased with the commisioning of a ''second-generation'' mosaic CCD camera with eight $2048\times4096$ pixel detectors in 2001. In the OGLE-III phase (2001 to 2009), 92 sq. degrees around the Galactic center were regularly observed. Currently, in the fourth phase of the project, OGLE-IV (since 2010), a 32 CCD detector, 260 Megapixel camera is being used. This allows for regular monitoring of over a billion sources, making OGLE-IV one of the largest sky variability surveys thus far. The OGLE-IV observations analyzed here cover 140 sq. degrees of the Galactic bulge.

The cadence of observations varies from 20 minutes to a few days. The majority of measurements were taken through the $I$-band filter, closely resembling that of the standard Johnson-Cousins system. The accuracy of the photometry is 0.02 mag for $I=17$ mag, 0.05 for $I=18.5$ mag, and 0.1 mag for $I=19.5$ mag. For details of the observing setup, reductions, and photometric and astrometric calibrations, we refer inquisitive readers to Udalski (2003), Udalski et al. (2008, 2015), and Szyma\'nski et al. (2011).

The time-series photometry of objects presented in this paper is available to the astronomical community from the OGLE Internet Archive.\footnote{ftp://ftp.astrouw.edu.pl/ogle/ogle4/NOVAE/BLG} The real-time photometry of future novae will be available on the webpage\footnote{http://ogle.astrouw.edu.pl/ogle4/transients} of the OGLE-IV Transient Detection System (Wyrzykowski et al. 2014).

\section{Nova eruptions}
\label{section:eruptions}

\subsection{A search for missing eruptions}
 
We found that 35 nova eruptions (known from the literature; for details see Section \ref{sec:searcholdnovae}) were observed by the OGLE survey in the Galactic bulge fields. However, it is known that the sample of Galactic novae is incomplete and some eruptions might have been missed. Indeed, in our initial search (Mr\'oz \& Udalski 2014), we found three novae from 2010-11 (OGLE-2010-NOVA-01, OGLE-2010-NOVA-02, OGLE-2011-NOVA-02) that had been overlooked.

We decided to extend our search to all of the light curves in the Galactic bulge databases (Section \ref{sec:data}). We found four additional, previously unknown novae. We name these novae: OGLE-2006-NOVA-01, OGLE-2008-NOVA-01, OGLE-2010-NOVA-03, and OGLE-2013-NOVA-04. All of the objects are listed in Table \ref{tab:novae}.

\subsection{Eruption light curves}

Nova eruptions have a large diversity of shapes and timescales. However, they can be generally described in terms of a ''universal decline law'' as broken power-law curves (Hachisu \& Kato 2006, 2015). A comparison with theoretical models can provide an estimate of a white dwarf mass and composition.

Additionally, a variety of secondary features may be superimposed on the monotonic nova decline. Strope et al. (2010) introduced a phenomenological classification. They defined seven light-curve classes: S (smooth), P (plateau), D (dust dip), C (cusp), O (oscillations), F (flat-topped), and J (jitter). See Strope et al. (2010) for definitions and prototypes.

In our sample of 35 novae with sufficient light-curve coverage, 23 are S type (66\%), three O (9\%), three D (9\%), two J (6\%), two P (6\%), and two C (6\%). For the classification of individual objects, see Table \ref{tab:novae_properties}. Because the early stages of some eruptions are missed, they could be misclassified. Two unusual novae, OGLE-2006-NOVA-01 (C class) and V5585 Sgr (J class), are described in detail in Sections \ref{sec:2006NOVA} and \ref{V5585Sgr}, respectively. The light curves of all eruptions are presented in Figure \ref{fig:lcs1}.

\subsection{Pre- vs. post-novae}
\label{section:prepost}

Collazzi et al. (2009) analyzed the pre- and post-eruption brightness of 30 CNe. They noticed that the majority of them did not change in the mean brightness after the eruption, implying that there is no change in accretion rate. However, they noted that five objects (V723 Cas, V1500 Cyg, V1974 Cyg, V4633 Sgr, and RW UMi) were significantly brighter after the nova outburst, suggesting that the explosion caused an increase in the mass-transfer rate. One of these objects (V4633 Sgr = Nova Sgr 1998) is still linearly fading with a rate of $91.9\pm16.4$ mmag yr$^{-1}$ in $I$ (note that Collazzi et al. 2009 report observations from 2009 suggesting that the decline has stopped, which -- at least in the $I$ band -- is not true). This system shows eclipses (see Section \ref{V4633Sgr}), and so, in principle, period changes can be measured. We were able to set an upper limit of $|\dot{P}_{\rm orb} | < 3\cdot 10^{-9}$ s s$^{-1}$. 

Among the objects described in this section, we successfully identified progenitors for 22 novae. For the next seven objects, pre-novae were fainter than the detection limit ($I \gtrsim 21$ mag). Additionally, we measured the mean brightness after the eruption, provided it did not exhibit systematic variations (a few slow novae are still declining, i.e. V2574 Oph, V5579 Sgr, V5582 Sgr, OGLE-2010-NOVA-02, V5592 Sgr). The mean difference $\delta I = I_{\rm pre} - I_{\rm post}$ is $0.33$ mag with a standard deviation of 0.36 mag. After neglecting four outliers (V5585 Sgr, V5587 Sgr, V5591 Sgr, and Nova WISE), we obtain $\langle\delta I\rangle= 0.21$ mag with a standard deviation of 0.26 mag. This result is in agreement with that of Collazzi et al. (2009),  who measured $\delta m=0.16$ mag with a standard deviation of 0.42 mag for non-RNe. We note that the variance of their result is larger because they had only single datapoints for several novae (from the Palomar Observatory Sky Survey plates). 

We found four stars that are significantly brighter a few years after the nova eruption than pre-novae. The slope of their light curves is practically zero, and so it is unlikely that they are still declining from eruption. Those exceptions are V5585 Sgr ($\delta I > 3.2$ mag), V5587 Sgr ($\delta I > 1.0$ mag), V5591 Sgr ($\delta I \approx 0.8$ mag), and Nova WISE ($\delta I \approx 0.6$ mag). Details can be found in Table \ref{tab:novae_properties}.

\subsubsection{Pre-eruption rises}

In his classic work, Robinson (1975) found that out of 11 novae with pre-eruption observations, 5 exhibited significant brightening preceding the outburst. However, Collazzi et al. (2009) revised these observations and analyzed more objects. They ruled out pre-eruption rises in some stars, but the progenitors of two novae -- V533 Her and V1500 Cyg -- did indeed brighten before the explosion. For a discussion of possible explanations, we refer the interested reader to Collazzi et al. (2009).

Out of 29 novae with pre-eruption data analyzed here, only 1 object -- V5587 Sgr -- showed significant pre-eruption brightening. In the months preceding the outbursts, the progenitor brightened by at least $\Delta I\approx 1$~mag with a mean rate of $-1.56\pm 0.06$ mag yr$^{-1}$.

\subsection{Orbital periods}

We used the analysis of variance method (Schwarzenberg-Czerny 1996) to find periodic variations in the light curves of all of the objects. In some cases, it was necessary to detrend them (by subtracting a cubic spline). For nine objects we found probable orbital periods. Their distribution has two peaks: between $0.15-0.35$ and $1.2-1.6$~days. These correspond to main-sequence and evolved secondaries, respectively. We checked that the distribution of orbital periods of our objects does not differ from the larger sample from Bode \& Evans (2008). Light curves folded with the orbital period are plotted in Figure \ref{fig:eclipsing_novae1}. We used quiescence data and, in some cases, we added data from late parts of eruption (after subtracting a cubic spline). All of the objects are described in detail below.

\begin{figure*}
\centering
\includegraphics[width=0.9\textheight,angle=90]{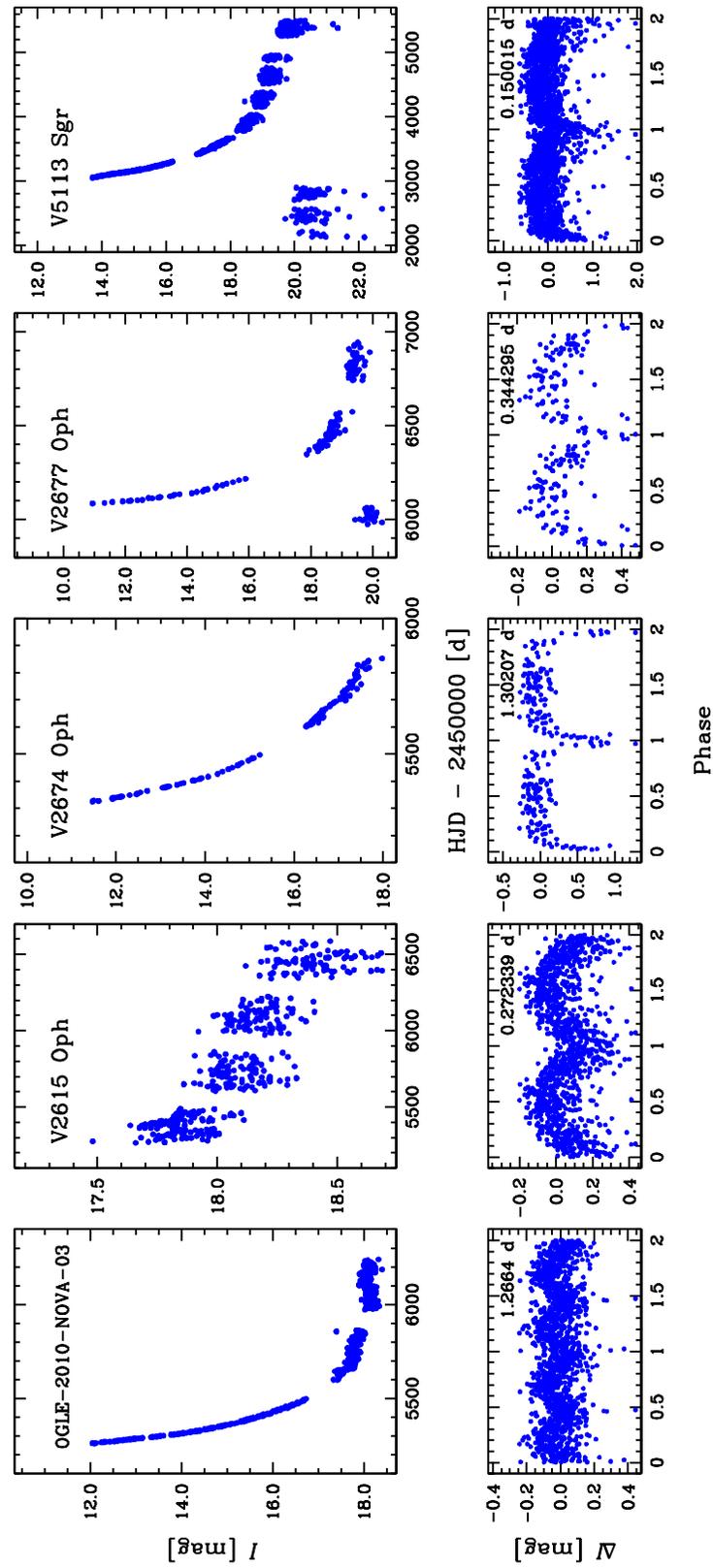}
\caption{Light curves of novae showing variations on an orbital period. Upper row: a close-up of the eruption. Lower row: detrended light curve folded with the orbital period. The value of period is in the upper right corner.}
\label{fig:eclipsing_novae1}
\end{figure*}

\addtocounter{figure}{-1}

\begin{figure*}
\centering
\includegraphics[width=0.72\textheight,angle=90]{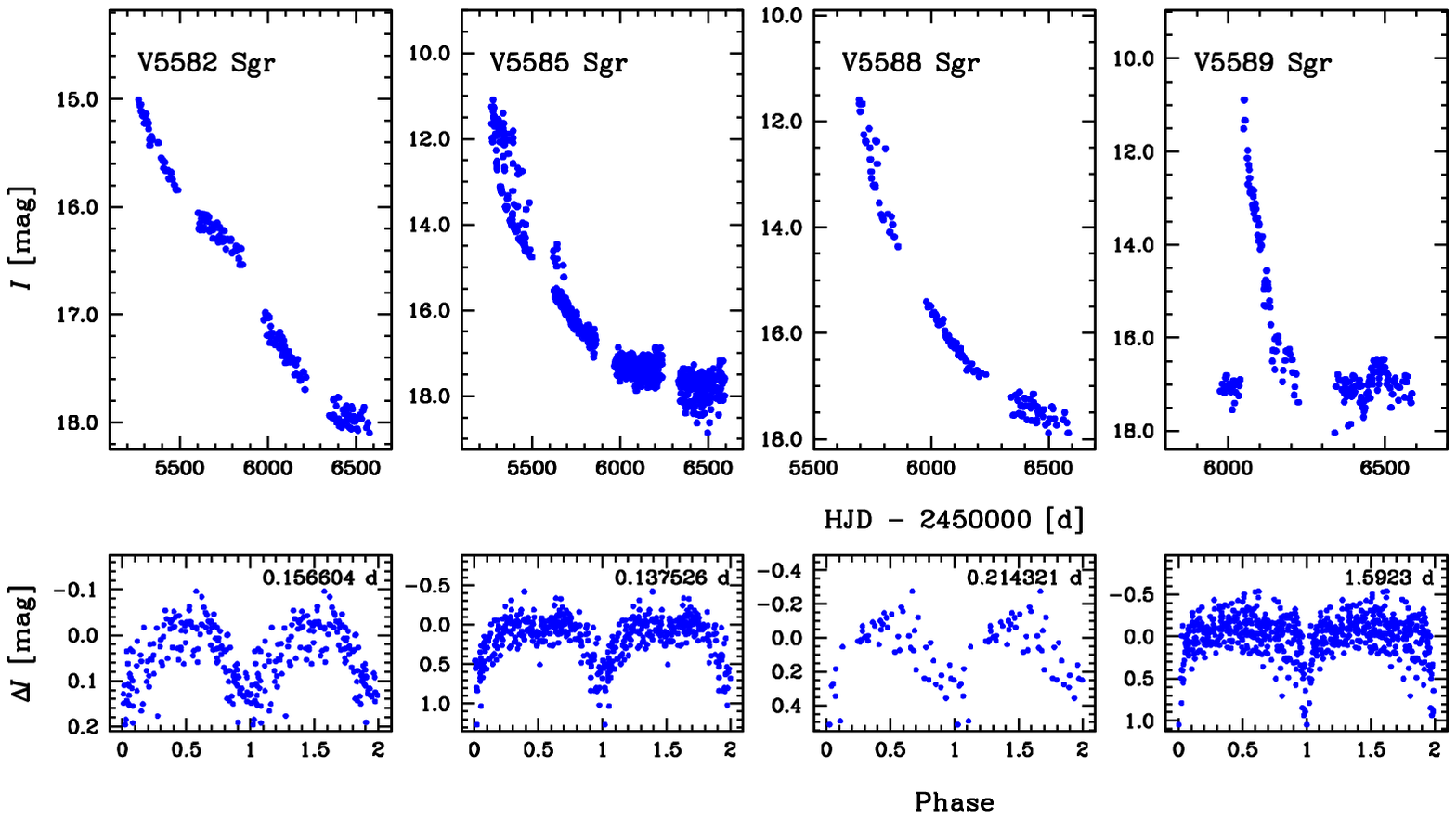}
\caption{(Continued)}
\end{figure*}

\subsection{Comments on individual objects}

\subsubsection{OGLE-2006-NOVA-01}
\label{sec:2006NOVA}

The eruption of this nova must have happened sometime during the seasonal gap between 2005 October and 2006 February. The first OGLE-III exposure in outburst was acquired on 2006 February 10.34622 UT. A few days later, the decline stopped for a month -- during which time, the brightness was practically constant at $I \sim 16.7$ mag. This was followed by a sudden drop by $\sim 1$ mag over less than 4 days (see Figures \ref{fig:cusp} and \ref{fig:lcs1}). Such a characteristic feature, called the ''cusp,'' indicates that this nova belongs to the rare C class (Strope et al. 2010; they list only three other novae with cusp-like secondary maxima, i.e. V1493 Aql, V2362 Cyg, and V2491 Cyg). Cusps might be caused by the release of additional magnetic energy from a strongly magnetized white dwarf, as proposed by Hachisu \& Kato (2009). A sudden brightness drop is probably caused by the formation of dust (Lynch et al. 2008).

Before the eruption in 2005 July -- August, we noticed a small brightening in the light curve, to $I\sim19.8$ mag. However, after a careful inspection of CCD images, we can rule out that this feature was real. The progenitor of the nova was below the magnitude limit. This false additional signal was most likely caused by somewhat bad weather conditions (high background, poor seeing). The remnant was not detected in the OGLE-IV frames (2010-on). We have not found any periodic signals in the light curve, even after detrending.

\begin{figure}
\centering
\includegraphics[width=0.35\textwidth]{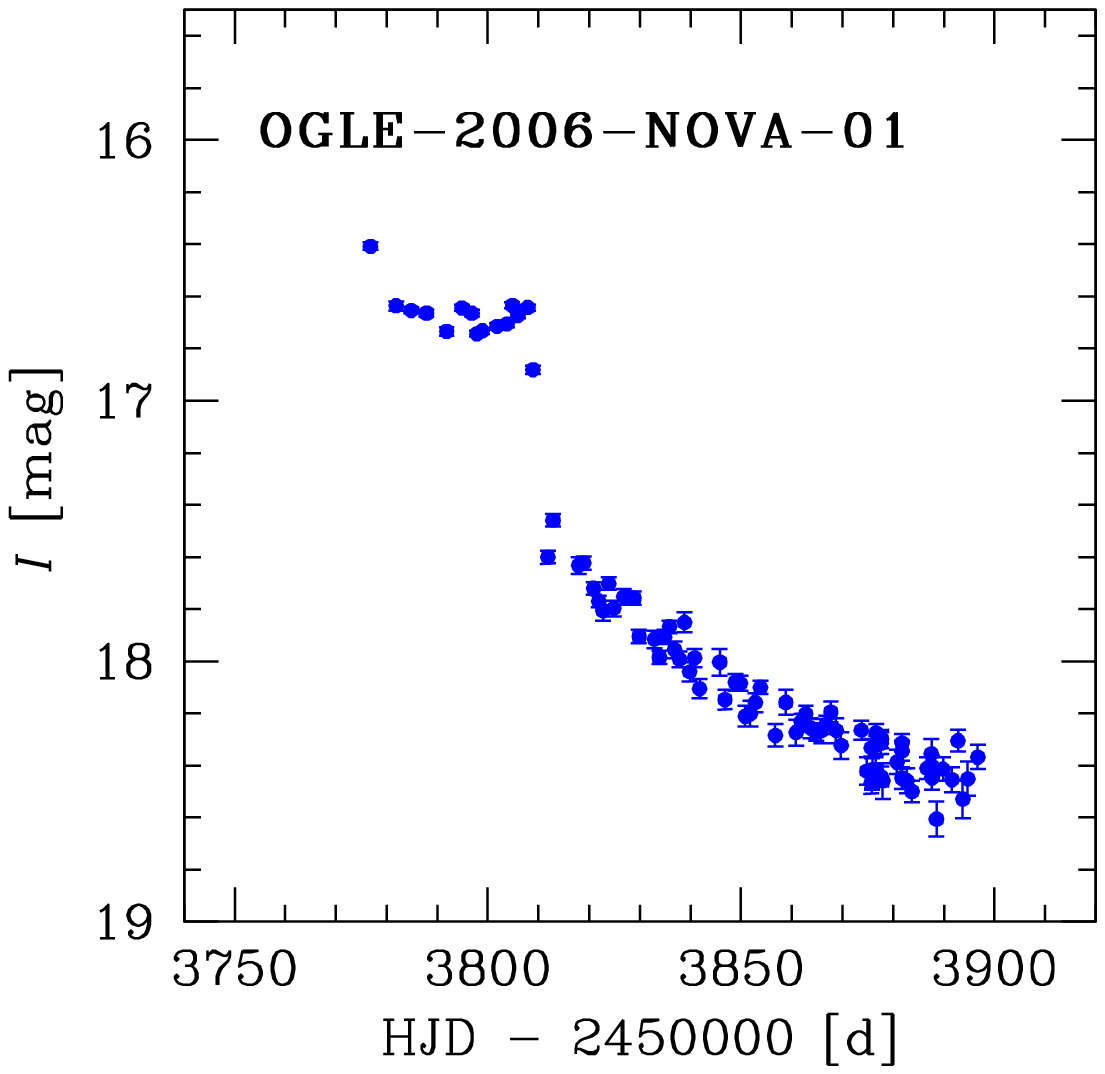}
\caption{Close-up of the cusp-like secondary maximum of OGLE-2006-NOVA-01.}
\label{fig:cusp}
\end{figure}

\subsubsection{OGLE-2008-NOVA-01}

This nova was captured in the first images from the 2008 observing season, on February 10.38723 UT. The eruption must have started between 2007 November and 2008 February, when the Sun passed in front of the Galactic bulge. This object was first spotted in OGLE data by T.~Mazewski (priv. comm.) and later independently revealed in our systematic search. The decline was smooth (an S class according to Strope et al. 2010). From the fading rate, we are able to set upper limits of $t_2<60$~days and $t_3<90$~days.\footnote{A speed of nova decline is parametrized by $t_n$ which is the time required for a decline by $n$ mag from the peak. Usually, $n=2$ or $n=3$ is used.} The nova returned to the pre-eruption state in late 2009.

\subsubsection{OGLE-2010-NOVA-03}

This is another nova which was discovered in the archival OGLE data and not detected in real time. It probably exploded in the first quarter of 2010. The first detection dates back to 2010 March 5.3843 UT. This was followed by a long, smooth decline (an S type in Strope et al. 2010 framework) on timescales of $t_2 \lesssim 55$~days and $t_3 \lesssim 82$~days.

After detrending the light curve, we found a strong sinusoidal signal at 0.63320(5)~days. This modulation is a signature of ellipsoidal variations from the secondary; the true orbital period is probably two times longer, i.e. $P_{\rm orb} = 1.2664(1)$~days (Figure \ref{fig:eclipsing_novae1}). After the eruption, the star is located redwards of the main sequence on the color-magnitude diagram (CMD). Both arguments show that the secondary is slightly evolved, possibly a subgiant.

\subsubsection{OGLE-2013-NOVA-04}

This star appeared in the ''new'' database of the BLG664.06 field (i.e., the object is invisible on the reference image) on 2013 March 31.34868 UT during the rise to maximum. The nova peaked at $I\approx 13.2$~mag on 2013 April 11. During the decline, we recorded a few low-amplitude oscillations (an O type according to Strope et al. 2010).

We measured decline timescales of $t_2 = 54$~days and $t_3 = 100$~days.

\subsubsection{V2615 Oph (Nova Oph 2007)}

This nova was discovered by Hideo Nishimura (2007) in images obtained on 2007 March 19.812 UT and spectroscopically confirmed on 2007 March 20.84 UT by Naito \& Narusawa (2007). Munari et al. (2008) estimated decline times in the $V$-band of $t_2 = 26.5$~days and $t_3=48.5$~days, a distance of $3.7 \pm 0.2$ kpc, and the interstellar reddening of $E(B-V)=0.90$ mag. Das, Banerjee \& Ashok (2009) detected carbon monoxide (CO) bands in infrared spectra, which enabled the study of the evolution and chemistry of a nova outflow during the early stages of eruption.

Our observations from 2010 -- 2013 showed that the nova has been declining with a rate of 0.122(6) mag yr$^{-1}$. After detrending the light curve, we measured a period of 0.272339(1)~days, which is likely the orbital period (see Figure \ref{fig:eclipsing_novae1}). The orbital period might be twice as long, but the former interpretation is supported by the location of the star on the CMD, bluewards of the main sequence.

\subsubsection{V2674 Oph (Nova Oph 2010b)}

This nova was discovered by Hideo Nishimura (2010) on frames taken on 2010 February 18.845 UT and confirmed spectroscopically by Imamura, Tanabe \& Fujii (2010). Munari, Dallaporta \& Ochner (2010) reported decline times of $t_2=18$~days and $t_3=31$~days in the $V$ band. They assessed the reddening as $E(B-V)=0.7 \pm 0.1$ and a distance to nova of $\approx 9$~kpc.

The eruption ended by the end of 2011. In the light curve of the post-nova, we found a periodic signal of 1.30207(6)~days, which is likely the orbital period (Figure \ref{fig:eclipsing_novae1}). If the period were doubled, then the primary and secondary eclipses would be exactly the same, which is unlikely.

\subsubsection{V2677 Oph (Nova Oph 2012b)}

The nova was discovered by John Seach in six exposures obtained on 2012 May 19.484 UT (Seach 2012). Walter et al. (2012) reports that this is a Fe II-type nova. The width of $H_{\alpha}$ is ${\rm FWHM} \approx 3000$ km s$^{-1}$.

The decline was smooth and ended in the late 2013. After detrending the light curve, we detected a period of 0.344295(8)~days, which is likely the orbital period (Figure \ref{fig:eclipsing_novae1}).

\subsubsection{V5113 Sgr (Nova Sgr 2003b)}

Nova Sgr 2003b was discovered by N. Brown from exposures taken on 2003 September 17.52 UT, when the star was $\sim 9.2$ mag (Brown 2003). Early spectra showed Balmer and Fe II emission lines with P-Cygni profiles, and hence the nova belongs to the Fe II class (Liller 2003; Ruch et al. 2003). The optical and infrared spectra obtained nine months after the eruption revealed the presence of coronal lines of [Si VI] and [S VIII] (Rudy et al. 2004). In the early light curve, up to two months after the eruption, several rebrightenings were observed (Tanaka et al. 2011). Tanaka et al. (2011) also noticed that P-Cygni profiles have disappeared between successive brightness maxima.

In the last image taken before the eruption (on 2003 September 13.10942 UT), the nova was fainter than $I>20$ mag. The moment of explosion has not been covered. The star was highly overexposed on all OGLE frames to the end of the 2003 observing season. Observations of the BLG101 field returned in 2004 February after the seasonal gap. The star was $I\sim 14$ mag, slowly and smoothly declining from the peak (see Figure \ref{fig:lcs1}). The decline has ended in 2010 (the remnant is $\delta I=0.5$ mag brighter than the pre-nova). In the light curve, we detected a strong periodic signal at 0.150015 days (see Figure \ref{fig:eclipsing_novae1}). We verified that this period is real and not an alias thanks to high-cadence ($\sim 20$ minutes sampling) photometric data.

\subsubsection{V5582 Sgr (Nova Sgr 2009b)}

The discovery of V5582 Sgr was reported by Guoyou Sun and Xing Gao (2009), who noticed a possible nova ($\approx 11.5$ mag) in images taken during the course of the Xingming Nova Survey on 2009 February 23.947-23.963 UT. A low-resolution spectrum, obtained on 2009 May 26.7 UT, showed strong emission lines of Balmer series, [O III], [N II], and He I (Kinugasa et al. 2009).

We noticed a faint ($I\approx 19.4$ mag) nova progenitor in OGLE-III images collected between 2001 and 2003. Observations of this star were resumed in 2010, when the nova was still declining from the maximum, suggesting rather large $t_3$. A plateau is visible in the data from early 2011. In 2013, the star was 1.5 mag brighter than the progenitor, while the decline slope had dropped practically to zero. This object may belong to a rare class of novae which substantially brighten  during the eruption.

After detrending the light curve, we detected a periodic signal of 0.156604(1)~days, relatively small amplitude $\Delta I\approx 0.2$ mag, and sinusoidal shape (see Figure \ref{fig:eclipsing_novae1}). Eclipses should be narrower, and so we suspect that the variability might be caused by an orbital hump from the hot spot. However, ellipsoidal modulations with a twice as long period (0.313208(2)~days) cannot be ruled out.

\subsubsection{V5585 Sgr (Nova Sgr 2010)}
\label{V5585Sgr}

This nova was discovered by John Seach in three exposures obtained on 2010 January 20.72 UT (Seach 2010), when it was $\sim 8.5$ mag. The exact moment of the eruption cannot be derived; it happened sometime between 2009 November 15.89 and 2010 January 20.72, when the nova was invisible due to its proximity in the sky to the Sun. The nova was spectroscopically confirmed by Maehara (2010). The spectra, taken on 2010 January 23.887 UT, showed emission lines of $H_{\alpha}$, $H_{\beta}$, and Fe II, which suggest that this object belongs to the ''Fe II'' class of novae. Two days after the discovery, Kiyota (2010) measured $V = 8.98$ mag, $V-R = 0.6$ mag, and $V-I=1.2$ mag.

This nova is poorly known. The American Association of Variable Star Observers (AAVSO) database contains only a couple of measurements, mainly in a ''green'' filter or visual observations. 

Our photometry began in the second half of 2010 March, when observations were resumed after starting the OGLE-IV phase. The nova was still bright, $I \sim 12$ mag, suggesting rather intermediate decline ($t_3\lesssim 35$ days). The full light curve reveals interesting ''rebrightenings'' or ''secondary maxima'' (Fig. \ref{fig:V5585Sgr} and \ref{fig:lcs1}), which were observed until 2011 May (for 16 months after the discovery). 

According to the Strope et al. (2010) classification scheme, V5585 Sgr is of the J (''jitter'') type. Objects from this class show short, sharp-peaked brightenings (flares) superimposed on the typical smooth, power-law decline. Jitters have amplitudes of at least 0.5 mag and are well-isolated in time. They may be present around the maximum brightness of a nova (like V723 Cas, V2468 Cyg) or -- rarely -- after the peak (like V4745 Sgr, V5588 Sgr). The light curve of V5585 Sgr resembles the latter ones.

We recorded 15 rebrightenings, but this is simply a lower limit. Some might have occurred before the start of the OGLE observations and a few were not observed during the seasonal gap 2010 November -- 2011 February. We changed magnitudes into flux and detrended the light curve (by subtracting a cubic spline). Subsequently, for each rebrightening, we measured its duration, $I$-band amplitude, and total emitted energy. 

First rebrightnenings are long, they last 12--22 days. They have a trapezoidal shape: the rise to the maximum is very fast (typically $\sim 1$ day), then the brightness decays linearly for several days, which is followed by a quick decline. The next maxima are shorter, approximately a few days long. Because of the insufficient time coverage, we cannot robustly determine their shape. However, it seems that the pattern of fast rise -- slow linear decline -- fast decline has not changed. The last rebrightnenings, observed in early 2011, were short-lived, $1-3$ days, and we have only a single datapoint available during each rebrightening. Nonetheless, there is a clear relation: the duration of the rebrightenings in V5585 Sgr decreases exponentially with time $\Delta T(t) \propto e^{-t/\tau_{\rm dur}}$ at a characteristic timescale of $\tau_{\rm dur}=124 \pm 12$ days.

Generally, the time-span between subsequent maxima decreases linearly with time: from 28--30 days in the early 2010 to 20 days at the end of that year. However, late rebrightenings are somewhat erratic; they did not emerge for a month between 2011 March and April. 

The amplitude of rebrightenings also changed with time. If we assume that the flux corresponding to $I=16$ mag is 1.0, then the early maxima peak around 40--60, while the last reach barely 2--5 units. The total energy emitted during the rebrightenings also follows an exponential pattern $\Delta E(t) \propto e^{-t/\tau_{\rm en}}$ with a characteristic timescale of $\tau_{\rm en}=72 \pm 5$ days.

The amount of data in the $V$ band is limited (see the color curve in Figure \ref{fig:kolorV5585Sgr}) but provides interesting clues concerning the nature of rebrightenings. Observations from 2011 show that the nova became redder during that year, from $V-I\approx-0.5$ to $V-I\approx 0.0$. However, during the first rebrightening, $V-I\approx 1.4$ mag was measured, while during the last one it is $V-I\approx 0.1$ mag. This indicates that the maximum of spectral energy distribution shifts from shorter to longer wavelengths during the rebrightenings. Note that Kiyota (2010) measured $V-I=1.2$ mag on 2010 January 22.862, soon after the discovery, and so oscillations might have started at that time or earlier.

In the late light curve from 2013, when the decline has ended, we detected eclipses. We measured an orbital period of 0.137526(2)~days. The light curve folded with this period is shown in Fig. \ref{fig:eclipsing_novae1}. The primary eclipse is deep, $\Delta I\sim 1 $ mag; secondary eclipses are shallower, $\sim 0.2$ mag. The ephemeris of primary eclipses is
\begin{equation}
{\rm HJD} = 2456335.0796(13) + 0.137526(2)\cdot E,
\end{equation}
while the upper limit for a period change is $|\dot{P}_{\rm orb}| < 1.2\cdot 10^{-7}$ s s$^{-1}$.

In the $V$-band frames collected prior to the eruption (in 2001--2003), we detected a faint progenitor $V=20.76$ mag. In the $I$-band exposures, it was invisible, $I>21$ mag, meaning that the amplitude of the eruption was substantial, at least 13 mag in $I$.

Finally, we would like to compare V5585 Sgr and two other well-observed novae that showed rebrightenings, i.e. V5588 Sgr (Munari et al. 2015; see also Figure \ref{fig:lcs1}) and V4745 Sgr (Strope et al. 2010). Munari et al. (2015) noticed that although V4745 Sgr and V5588 Sgr have similar light-curve shapes, their spectroscopic evolution was different. V5588 Sgr was a hybrid nova, which means that spectral features typical for the ''Fe  II'' and ''He/N'' class existed simultaneously, while V4745 Sgr did not develop He/N spectrum at all. During rebrightenings, emission lines in the spectrum of V4745 Sgr showed additional strong P-Cygni profiles; on the other hand, the spectrum of V5588 Sgr practically did not change. Unfortunately, there are no spectral data of V5585 Sgr published in the literature, and so we will limit ourselves only to photometry. 

The energy emitted during rebrightenings, both in V5585 Sgr and V5588 Sgr, decays exponentially with time. We measured a characteristic timescale of decay of $66\pm 9$~days from the data presented by Munari et al. (2015). A similar value (72~days) was found here for V5585 Sgr. 

Time intervals between successive rebrightenings increased with time, from 18 to 42 days in V5588 Sgr (Munari et al. 2015) and from 8 to 60 days in V4745 Sgr (Strope et al. 2010). Tanaka et al. (2011) noticed a similar systematic trend for a few other objects with secondary maxima. However, in the case of V5585 Sgr, time intervals shortened from 30 to 20 days; during the late decline, jitters were erratic. The duration of rebrightenings increased from 2.1 to 8.6 days for V5588 Sgr (it was quantified by the peak FWHM). Here, it exponentially decayed from 22 days to 1 day, while in V4745 Sgr there is no significant trend.

Some theoretical models by Hillman et al. (2014) can reproduce rebrightenings (see their model 065.30.10 and Figure 9). Although their shape resembles the light curve of V5585 Sgr, they have a much longer timescale (100--150 days). Those secondary outbursts may be produced by white dwarf envelope pulsations. Following mass-loss, the envelope contracts, leading to an increase in the pressure and nuclear luminosity. The heated envelope re-expands, producing successive secondary eruptions. We note that in this scenario, the nova should be blue in maxima, as opposed to what was observed in V5585 Sgr. 

\begin{figure}
\centering
\includegraphics[width=0.5\textwidth]{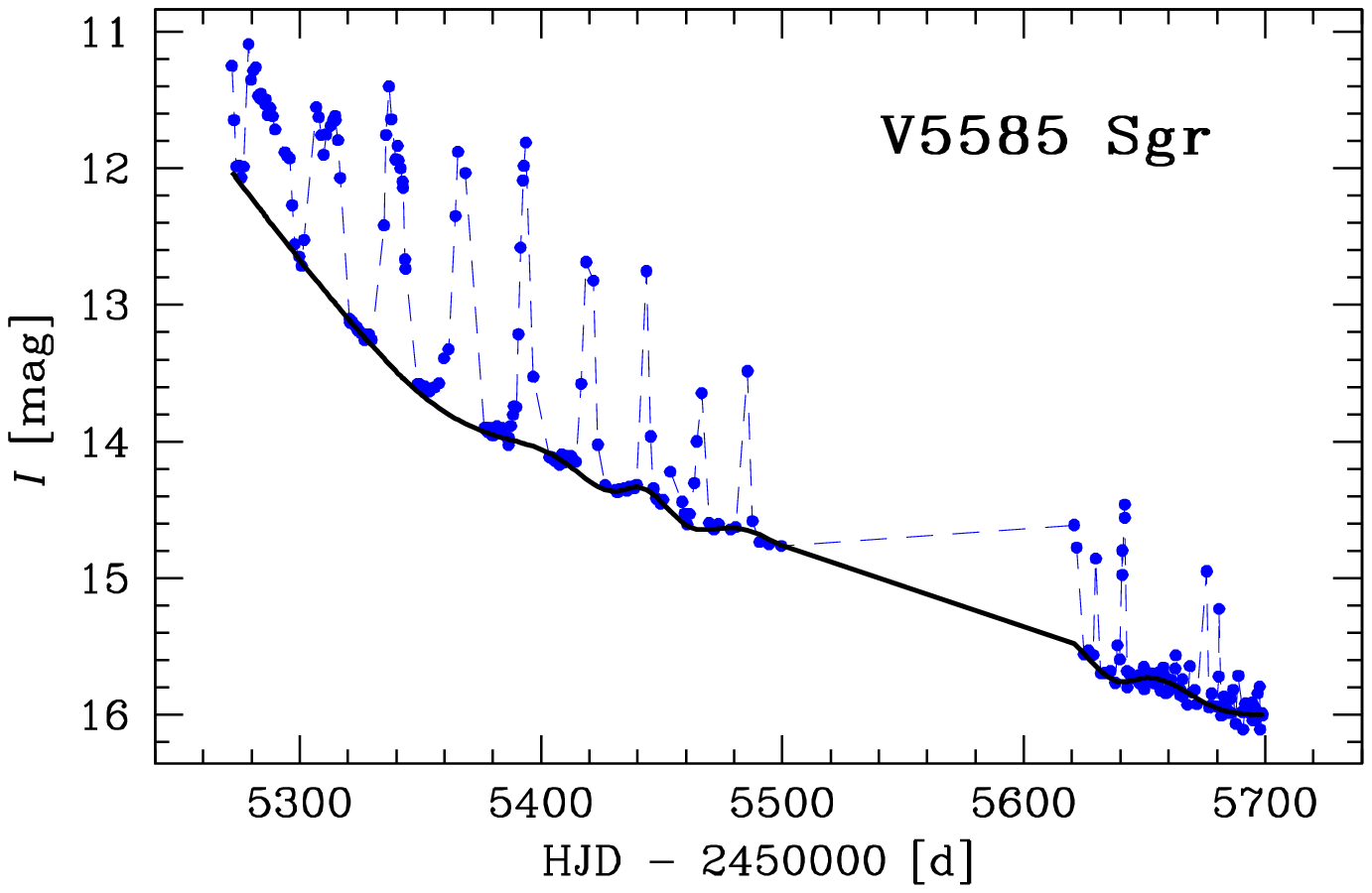}
\caption{Close-up of secondary maxima of V5585 Sgr (Nova Sgr 2010). Consecutive points are connected with the dashed line to guide the eye. The thick solid line is drawn from cubic spline interpolation and represents the underlying typical, smooth nova decline. }
\label{fig:V5585Sgr}
\end{figure}

\begin{figure}
\centering
\includegraphics[width=0.5\textwidth]{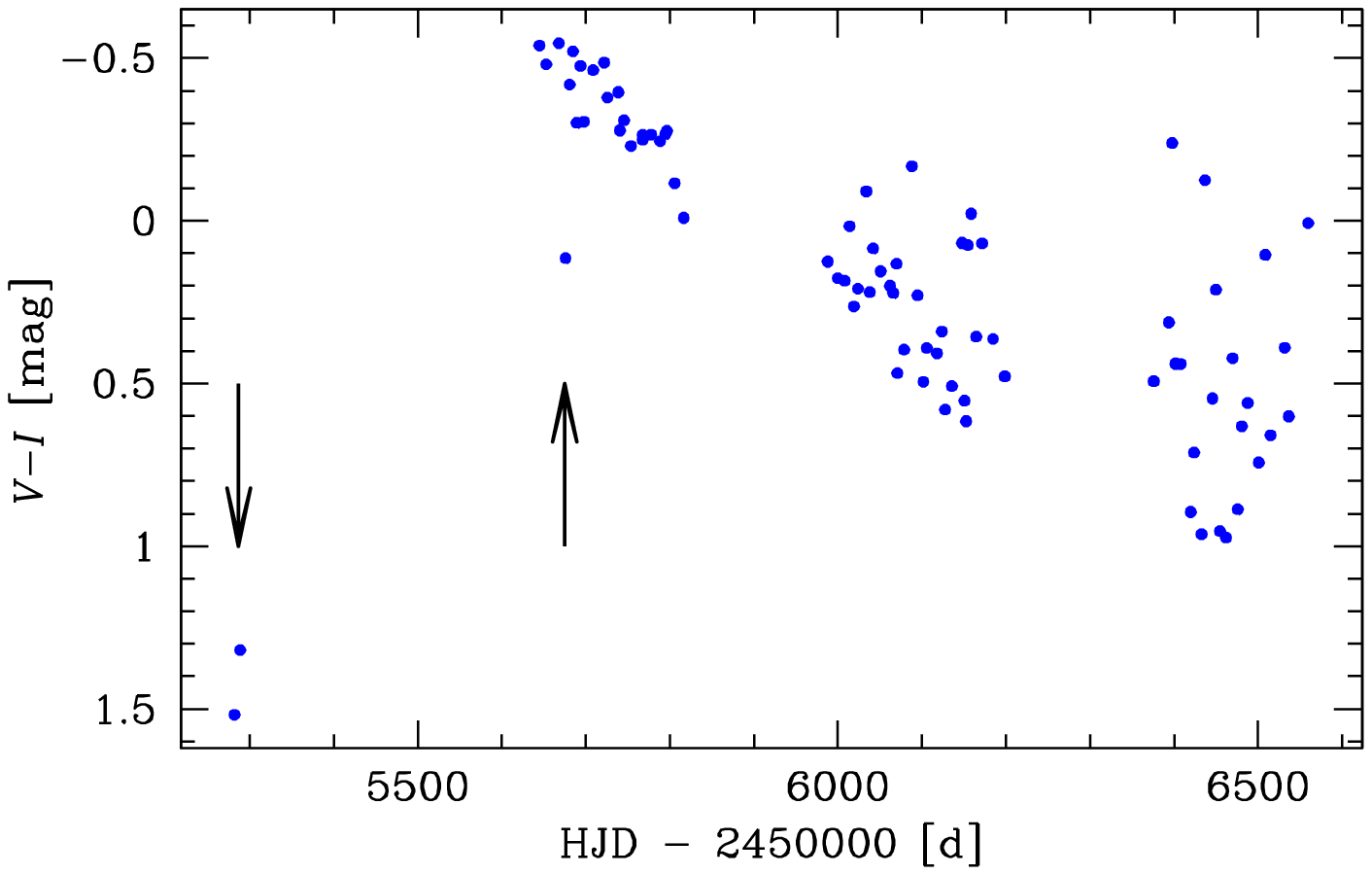}
\caption{Color curve of V5585 Sgr (Nova Sgr 2010). Arrows denote moments of rebrightenings during which the nova is substantially redder.}
\label{fig:kolorV5585Sgr}
\end{figure}

\subsubsection{V5588 Sgr (Nova Sgr 2011b)}

This object was discovered by Koichi Nishiyama and Fujio Kabashima (2011) in images taken on 2011 March 27.832 UT. V5588 Sgr belongs to a rare class of hybrid novae - it showed spectral features typical of Fe II and He/N types simultaneously (Munari et al. 2015). 

This nova exhibited six secondary maxima (rebrightenings; see Section \ref{V5585Sgr}), which are visible in our photometry (Figure \ref{fig:lcs1}). For a detailed description, we refer the interested reader to Munari et al. (2015). 

After detrending the light curve, we found a period of 0.214321~days (Figure \ref{fig:eclipsing_novae1}). 

\subsubsection{V5589 Sgr (Nova Sgr 2012)}

The nova was detected by Stanislav Korotkiy and Kirill Sokolovsky in images obtained on 2012 April 21.011 UT (Korotkiy \& Sokolovsky 2012). It peaked around $V\sim 9.0$ mag. The decline was very fast with $t_2 = 4.5\pm 1.5$~days and $t_3 \approx 7$~days according to Walter et al. (2012). 

Walter et al. (2012) also reports that this is a hybrid nova: it showed Fe II emission lines three days after the discovery, but five days later the spectrum was typical of He/N nova. Coronal emission lines of [Fe X], [Fe XI], and [Fe XIV] were observed 65 days after the eruption. The $H_{\alpha}$ emission line was broad with ${\rm FWHM} \approx 5700$ km s$^{-1}$.

In quiescence, we measured a strong periodic signal at $P = 1.59230(5)$~days. Eclipses would be exactly the same if this period were doubled, so this is likely the orbital period (Figure \ref{fig:eclipsing_novae1}). The long orbital period suggests that the secondary has evolved off the main sequence and is probably a subgiant. The amplitude of eruption ($\sim 10$ mag in $V$), fast decline, long orbital period, and broad $H_{\alpha}$ are indicators of a massive white dwarf and high mass-transfer rate (Pagnotta \& Schaefer 2014). That is why this system is a good RN candidate.

\section{Post-novae}
\label{sec:old}

Our knowledge of nova systems long after (or before) eruption is still far from complete, mainly because of a lack of observational data. For example, only $\sim 1/3$ of pre-1980 novae has been identified in quiescence (Tappert et al. 2013c). Studies of post-novae, particularly eclipsing ones, can provide a direct verification of evolutionary scenarios (e.g., hibernation -- Shara et al. 1986). Another open question is whether white dwarfs in nova binaries grow in mass to eventually explode as a SNe Ia?

There are some recent studies regarding post-novae. Saito et al. (2013) published a number of possible infrared counterparts from the Vista Variables in the Via Lactea survey (VVV; Minniti et al. 2010). 
In a series of papers, Tappert et al. (2012, 2013a,b, 2014) presented their successful observational program of identification and examination of post-novae. They use multicolor photometry to select candidates, confirm them spectroscopically, and finally carry out in-depth studies.

In this paper, we would like to present another approach. Benefiting from the long-term, superb-quality photometry collected over the course of the OGLE project, we search for variability of post-novae. We analyze variations occurring on orbital timescale (i.e., eclipses, superhumps, white dwarf rotation), as well as pulsations of evolved secondaries, dimming events, dwarf nova outbursts, and RN eruptions.

\subsection{A search for old novae}
\label{sec:searcholdnovae}

Nova eruptions have been recorded systematically since the second half of the 19th century. We found that the most complete list of historical novae (almost 450 objects) is that of the International Variable Star Index (VSX) database\footnote{http://www.aavso.org/vsx/index.php} (Watson 2006). Other compilations include the IAU Central Bureau for Astronomical Telegrams' list\footnote{http://www.cbat.eps.harvard.edu/nova\_list.html} (entries end in 2010), while Koji Mukai's list\footnote{http://asd.gsfc.nasa.gov/Koji.Mukai/novae/novae.html} contains the most recent discoveries.

We retrieved the coordinates (for the epoch J2000.0) of all known novae from the SIMBAD database.\footnote{http://simbad.u-strasbg.fr/simbad/} For 22 objects\footnote{V2290 Oph, FL Sgr, FM Sgr, V787 Sgr, V927 Sgr, V928 Sgr, V1012 Sgr, V1014 Sgr, V1015 Sgr, V1149 Sgr, V1150 Sgr, V1274 Sgr, V1583 Sgr, V1944 Sgr, V2415 Sgr, V4157 Sgr, V5586 Sgr, V719 Sco, V720 Sco, V722 Sco, V723 Sco, V902 Sco} more precise astrometry was available in VSX. According to the VSX webpage, it was taken from the General Catalog of Variable Stars (GCVS, version of 2009 November). 

We searched for counterparts in the OGLE databases within $3''$ around catalog positions. We inspected the light curves of all candidates by ''eye'' looking for eruptions, dimming events, pulsations, and all other suspicious variations. In this step, we also recovered 35 nova eruptions (see Section \ref{section:eruptions} for details).

Subsequently, we calculated periodograms (in a period range of 0.05--10 days, the analysis of variance method was used; Schwarzenberg-Czerny 1996). For nine objects, we found eclipses and ultra-short orbital periods, typically 2--5~hr. We believe that these are likely post-novae; they are listed in Table \ref{tab:ecl} and described in Section \ref{section:eclipsing}.

For the remaining novae, we chose the closest star to catalog positions from the OGLE databases. Among these, 1/3 are located closer than $0\farcs2$, 1/2 closer than $0\farcs5$, and 4/5 closer than $1\farcs0$. The mean number density of stars in the Galactic bulge fields varies from 200 to 2000 stars/arcmin$^2$, and so the expected number of stars in a $1''$ circle is 0.15--1.5. Therefore, it is probable that stars closer than $1''$ from the catalog position are in fact real nova counterparts. We call them ''possible'' post-novae. However, a definitive classification requires spectroscopy and it is outside the scope of this paper.

In the case of nine objects,\footnote{V908 Oph, V2024 Oph, V1174 Sgr, V1175 Sgr, V1572 Sgr, Nova Sgr 1963 (NSV 9828), Nova Sgr 1953 (NSV 10158), Nova Sco 1952 (NSV 9663), and Nova Sco 1954 (NSV 9808)} the accuracy of the astrometry was very poor (up to $0\farcm1$), and so we decided to extend the search radius to $10''$. At an angular distance of $7\farcs0$ from the catalog position of V1174 Sgr (Nova Sgr 1952b), we found star BLG512.14.146226 with a period of 0.3090452(4)~days. Its light curve is shown in Figure \ref{fig:eclipsing_post_novae}. For the remaining eight stars, we did not find any reliable variable counterparts. We neglected those stars in further analyses.

For all objects, we give the mean brightness in the $I$ and $V$ band (if available) from OGLE-IV or OGLE-III photometric maps (Szyma\'nski et al. 2011).  Depending on the position in the CMD, we marked ''MS'' those stars located on or bluewards of the main sequence and ''RG'' those on the red giant branch (RGB).

Finally, we end up with 9 likely (Table \ref{tab:ecl}) and 73 possible post-novae (Table \ref{tab:postnowe1}). In Tables \ref{tab:ecl} and \ref{tab:postnowe1}, we give designation, coordinates, distance from the SIMBAD position, colors, and types of the secondaries (if available). All of the objects are listed in lexicographical order.

\subsection{Variability types}

\subsubsection{Semi-regular variability}

All stars on the RGB and asymptotic giant branch show variability caused by oscillations. As the star evolves, its amplitude grows - from OGLE small amplitude red giants, semi-regular variables, to Miras (Soszy\'nski et al. 2013). Red giants can be members of a cataclysmic system and a quiescent light curve might be dominated by pulsations, especially at long wavelengths. Well-known examples are RNe V3890 Sgr and V745 Sco (Mr\'oz et al. 2014).

It is no accident that red giant secondaries are connected with RNe -- strong winds can drive the high mass-transfer rate. However, the presence of a red giant does not suffice for the RN -- additionally, a white dwarf must be massive. 

In Table \ref{tab:SRVS}, we list stars showing semi-regular variability. For completeness, we added RN V745 Sco. We emphasize once again that follow-up spectroscopy can prove that these are real post-novae. 

\subsubsection{Dwarf nova outbursts}

In the case of high mass-transfer rate systems with fully ionized, stable accretion disks, there are no outbursts. The vast majority of post-novae belong to this class of cataclysmic stars, and hence the name of the class is ''novalike variables.'' If the mass-transfer rate is lower, then so-called thermal-tidal instabilities in the accretion disk might occur, resulting in dwarf nova outbursts (Hellier 2001; Warner 2003). In fact, the hibernation scenario (Shara et al. 1986) predicts that decades after the nova eruption, the mass-transfer rate will gradually decrease (or even eventually cease) and dwarf nova outbursts may be observed in the system.

Only a few novae showed such outbursts. The best examples are V1017~Sgr (Webbink et al. 1987), GK~Per (Sabbadin \& Bianchini 1983), and V446 Her (Honeycutt et al. 1995). V1017 Sgr and GK Per are somewhat atypical (Sekiguchi 1992)  because they have long orbital periods, 5.7 and 1.99 days, respectively. A large orbital separation enables faster cooling of the secondary and the mass transfer decreases faster than in short-period systems.

The objects listed in Tables \ref{tab:ecl} and \ref{tab:postnowe1} have not exhibited dwarf nova eruption, demonstrating the rarity of this phenomenon. The oldest post-novae in our sample exploded in the early 20th century, meaning that for the majority of novae $\sim 100$ years is not enough for significant changes in mass-transfer rate.

However, we have recorded outburst of the star OGLE-2004-BLG-081, which might be a dwarf nova outburst in a post-nova system. This star was alerted as a candidate microlensing event by OGLE and independently by the MOA (Microlensing Observations in Astrophysics) group as MOA 2004-BLG-3. However, the shape of the outburst (see Figure \ref{fig:DNinCN}) does not fit the microlensing event, even after taking into account secondary effects (parallax, blending, binarity; Wyrzykowski et al. 2006). The outburst lasted $\approx 100$~days and had an amplitude of $\Delta I=2$ mag. The light curve is almost symmetric; however, on the rising branch, we noticed a short plateau (lasting 12 days). The outburst properties do not resemble those of typical dwarf novae, but the similarity to the outbursts of V1017 Sgr and GK Per is strikingly evident (see Webbink et al. 1987 and Sabbadin \& Bianchini 1983 for comparison).

A quiescent light curve of OGLE-2004-BLG-081, typical for a contact binary, shows eclipses with a period of 3.96659~days (Figure \ref{fig:DNinCNq}). Assuming that the secondary fills its Roche lobe, its density is 0.012 g cm$^{-3}$, suggesting that it is a subgiant. This is consistent with the location on the CMD. 

The star follows the relation between outburst duration and orbital period for dwarf novae. According to Warner (2003), the outburst width $\Delta T_{0.5}$ (within 0.5 mag of maximum) is correlated with the orbital period $P_{\rm orb}$: $\log \Delta T_{0.5} = 0.80\log P_{\rm orb}-0.05$, while the decay time $\tau_{\rm d}$ follows $\log\tau_{\rm d} = 0.84\log P_{\rm orb}-0.28$. ($P_{\rm orb}$ is expressed in hr, $T_{0.5}$ -- in days, $\tau_{\rm d}$ -- in days~mag$^{-1}$). The accuracy of these relations is around 20\%. For $P_{\rm orb} \approx 95.2$ hr, we estimate $T_{0.5}\approx34 \pm 7$~days and $\tau_{\rm d}\approx 26 \pm 6$ days~mag$^{-1}$, which agree with the observed outburst parameters: $T_{0.5}^{\rm obs}=43\pm 5$~days and $\tau_{\rm d}^{\rm obs}\approx 29.9 \pm 0.6$ days~mag$^{-1}$. 

Sartore \& Treves (2012) found that this star matches the X-ray source 2XMM J180540.5-273427. They speculate that this object is the isolated black hole which caused the microlensing event in 2004. We are sure this is not the case. 

The shape of the outburst of OGLE-2004-BLG-081 and its orbital period indicate a similarity to V1017 Sgr and GK Per. We suspect that it was a dwarf nova outburst in a post-nova system. This hypothesis would be strongly supported by the detection of the nova shell around this star. 

\begin{figure}
 \includegraphics[width=0.5\textwidth]{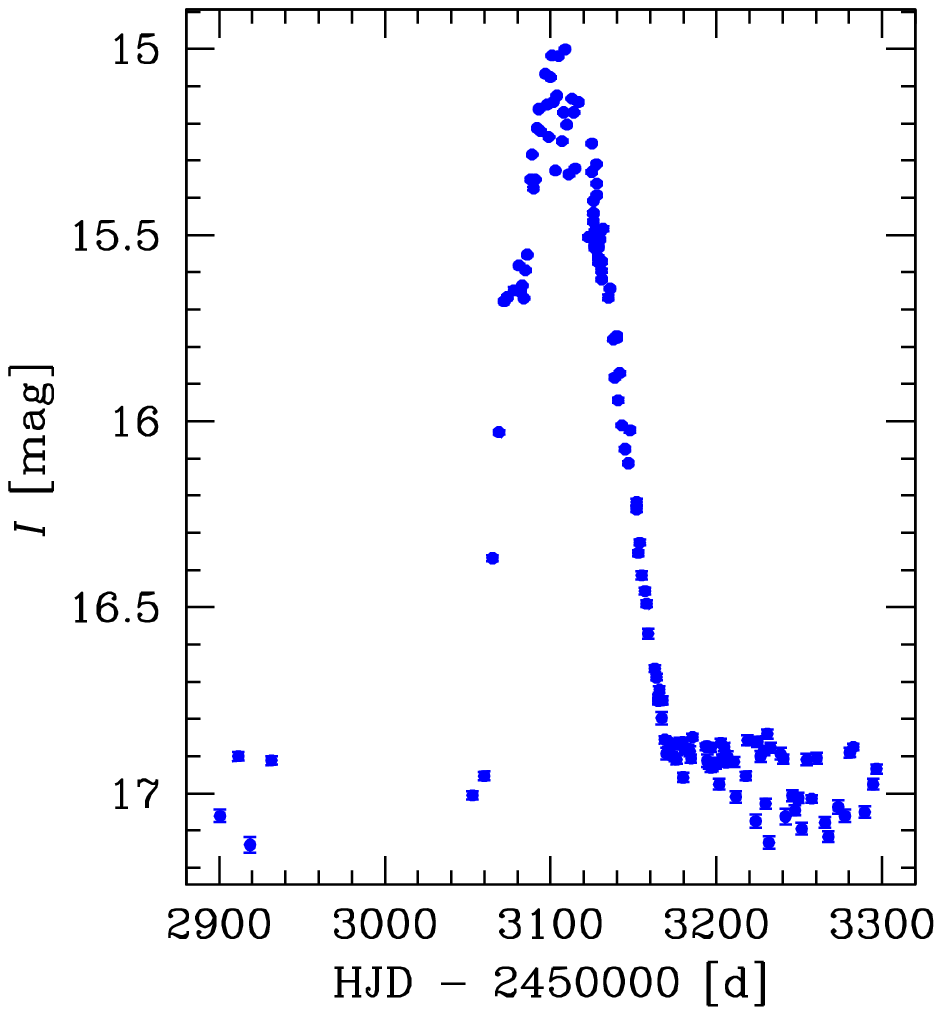}
 \caption{Close-up of the 2004 eruption of OGLE-2004-BLG-081. Note the similarity to the outbursts of V1017 Sgr and GK Per.}
 \label{fig:DNinCN}
\end{figure}

\begin{figure}
 \includegraphics[width=0.5\textwidth]{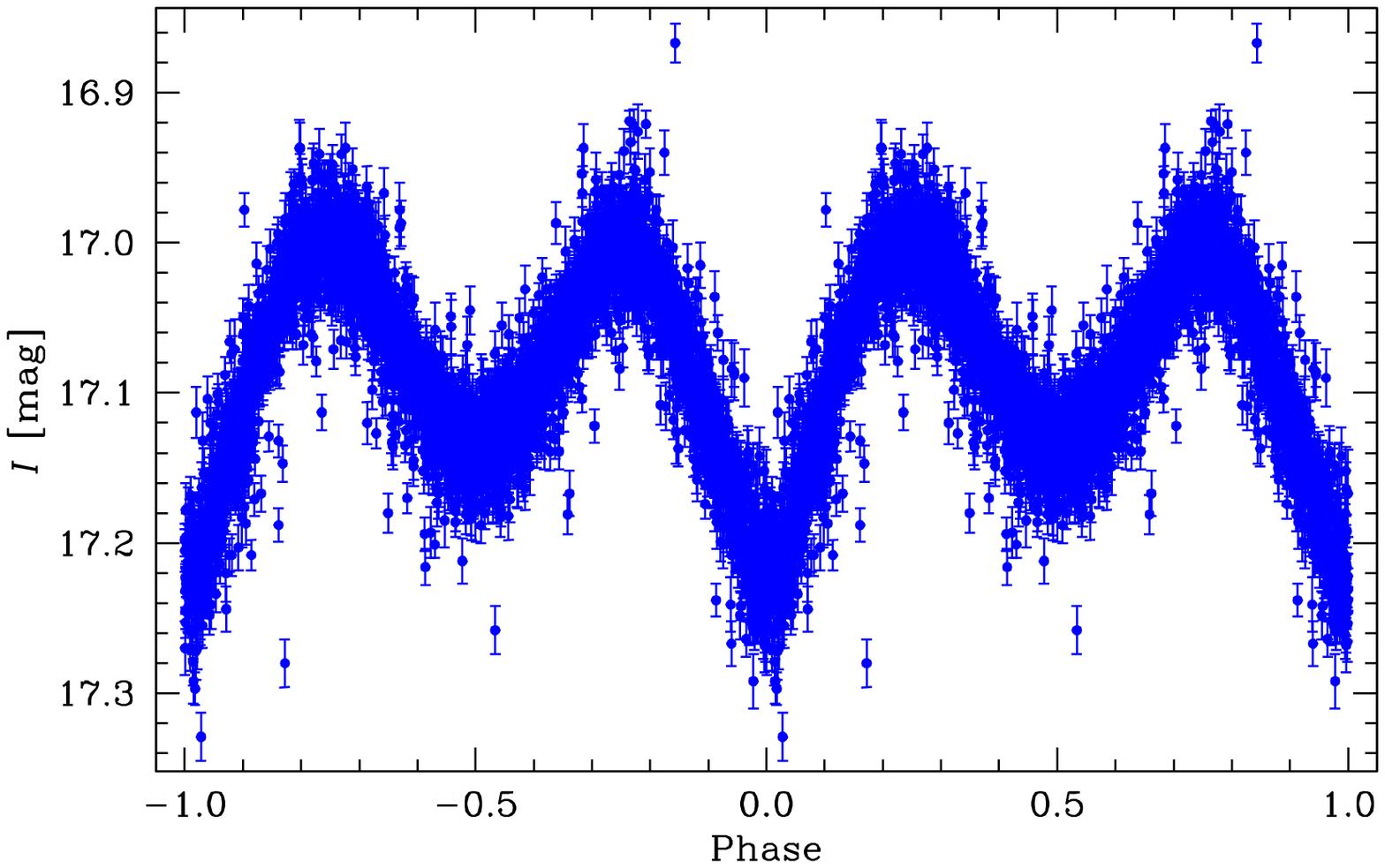}
 \caption{Quiescent light curve of OGLE-2004-BLG-081 phased with an orbital period of $P_{\rm orb}=3.96659$~days.}
 \label{fig:DNinCNq}
\end{figure}

\subsubsection{Recurrent novae eruptions}

None of the described objects showed the second nova eruption. We quantify this in Section \ref{section:howmanyRN}. Mr\'oz et al. (2014) has already described RNe in the OGLE fields.

Pagnotta \& Schaefer (2014) recently proposed several RN candidates, e.g., V4643~Sgr. Our photometry of this object is very limited, however, it shows either semi-regular pulsations or ellipsoidal modulation, implying a long orbital period.

\begin{figure*}
\centering
\includegraphics[width=0.7\textwidth]{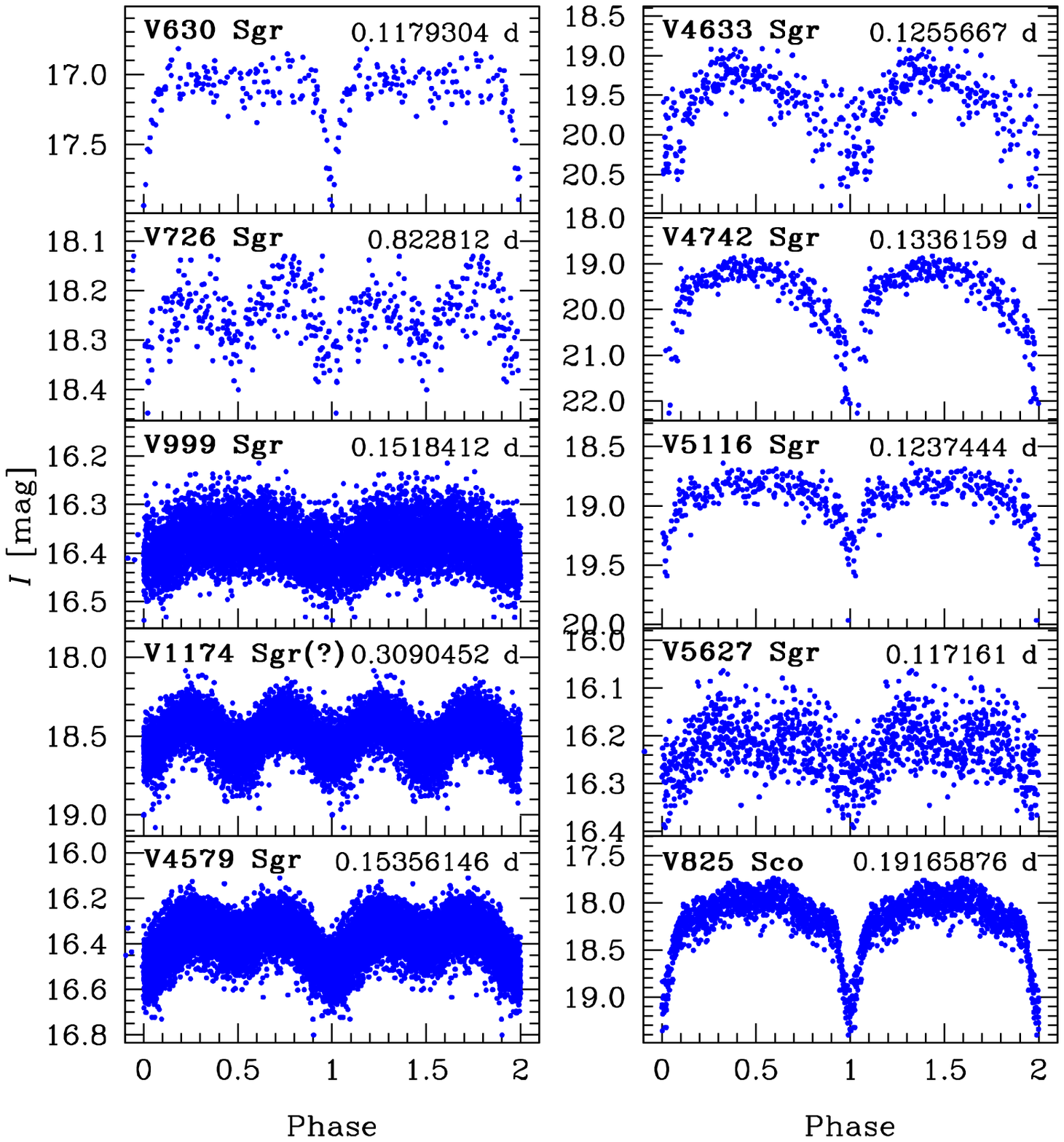}
\caption{Light curves of likely post-novae showing variations on an orbital period. The value of the period is in the upper right corner. }
\label{fig:eclipsing_post_novae}
\end{figure*}

\begin{figure*}
\centering
\includegraphics[width=0.5\textwidth]{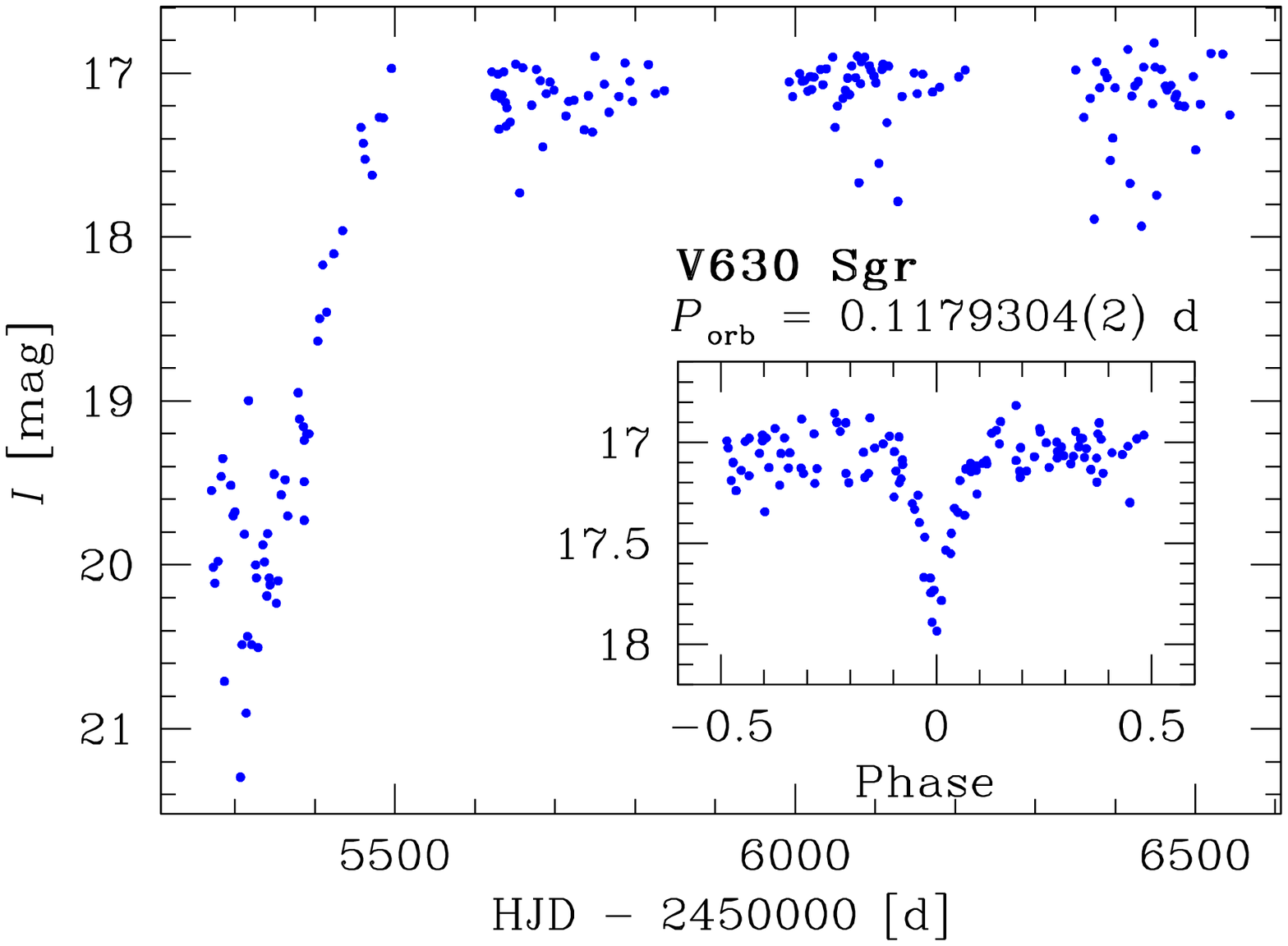}
\caption{Light curve of V630 Sgr (Nova Sgr 1936c) resembling that of a VY Scl-type star. Note the brightness drop in 2010.}
\label{fig:V630Sgr}
\end{figure*}

\subsection{Comments on individual objects}
\label{section:eclipsing}

\subsubsection{V630 Sgr (Nova Sgr 1936c)}

V630 Sgr was discovered by Sigeki Okabayasi in 1936 October at $V\approx 4.5$ mag (Okabayasi 1936). The nova is among the fastest known with a decline timescale of $t_3 = 6$ days (Schmidtobreick et al. 2005). The decline speed implies a low mass of ejecta, and hence a massive white dwarf. 

In their high-speed photometry, Woudt \& Warner (2001) noticed shallow eclipses and derived an orbital period of 0.1180~days. Fourier analysis showed the presence of the secondary periodicity at 0.1242~days, 5.3\% longer, which they interpret as superhumps. The spectrum of the post-nova -- typical of a CV -- is dominated by Balmer and He II lines in emission (Schmidtobreick et al. 2005).

The strongest periodicity in the OGLE light curve is at 0.1179304(2)~days, in agreement with Woudt \& Warner (2001). We show data folded with this period in Fig. \ref{fig:eclipsing_post_novae} and in the inset of Fig. \ref{fig:V630Sgr}. In the $I$ band, eclipses are somewhat deeper than in $V$, $\sim1 $ mag. 

We searched for additional periods by fitting and subtracting a high-order trigonometric polynomial from the data. In such a pre-whitened data set we did not find any significant periodicity (in a range of 0.05-0.5~days). However, we cannot completely rule out that V630 Sgr is a permanent superhumper as Woudt \& Warner (2001) suggest. Possibly, the number and cadence of observations in this case is too poor for a robust detection of low-amplitude superhumps.

An $O-C$ analysis yields an upper limit on period change of $|\dot{P}_{\rm{orb}}|<3\cdot 10^{-9}$ during three years of observations.

Woudt \& Warner (2001) collected their observations in 2001 August when V630 Sgr was around $V\approx 17.6$ mag. Our photometry starts in 2010, when the star was surprisingly faint, $I\sim 20$ mag (see Figure \ref{fig:V630Sgr}). In some frames it was almost invisible, close to the detection limit. In 2010 June, it started to brighten at a constant rate of $-6.7 \pm 0.3$ mag yr$^{-1}$, reaching 17 mag at the end of 2010. The amplitude was at least 3.5 mag, meaning that the star brightened at least 25 times.

This is very likely a VY Scl-type star. At maximum (''bright state''), VY Scl stars are ordinary nova-like variables with a high mass-transfer rate. In their low state, which may last up to a few hundred days, they drop by several magnitudes because the mass transfer drops or even switches off. A popular explanation invokes the passage of starspots across the Lagrangian point (e.g., Leach et al. 1999). 

Studies of eclipses in V630 Sgr in bright and low states might help to find the best explanation for the VY Scl phenomenon.

\subsubsection{V726 Sgr (Nova Sgr 1936b)}

This star was discovered by Luyten (1937) and independently by Mayall (1938). The nova exploded in the early May of 1936, peaking at $V\sim 10.5$ mag. It faded below the detection limit in late August with decline timescales of $t_2=45$ days and $t_3=95$ days. According to Duerbeck (1987), there have been several post-nova candidates. Proper identification is possible now with the OGLE data.

We have found a star showing variability with a period of 0.411406(1)~days, 0$\farcs$31 from the catalog position of V726 Sgr. Possibly, the orbital period is twice as long, i.e., 0.822812(3)~days and the light curve is produced by ellipsoidal variations of the secondary (see Fig. \ref{fig:eclipsing_post_novae}). 

Three pieces of observational evidence suggest that this star has an evolved secondary, possibly a subgiant: infrared colors of $(J-H)=0.74$ mag, $(H-K_s)=0.13$ mag (Saito et al. 2013), the location in the CMD, and the long orbital period. For this reason, the star meets some of the criteria for an RN (Pagnotta \& Schaefer 2014). However, the 1936 eruption was relatively slow -- all known RNe are fast or very fast (typically $t_3<15$~days; with exceptions of CI~Aql (32~days), T~Pyx (62~days), and IM~Nor (80~days); Schaefer 2010), which makes this hypothesis unlikely. Nevertheless, this system is worth follow-up studies.

\subsubsection{V999 Sgr (Nova Sgr 1910)}

This nova was immortalized on 16 photographic plates taken between 1910 March 21 and June 10 and discovered later by Fleming (1910). It peaked around 8 mag. The recorded spectrum showed prominent Balmer lines. The nova was slow ($t_3=160$ days; Duerbeck 1987) and some fluctuations were observed during the decline (Walker 1923). 

The remnant is relatively bright ($I\approx 16.4$ mag) and spectroscopic confirmation was possible. Duerbeck \& Seitter (1987) reported the spectrum to be dominated by a blue continuum with weak H$\alpha$ emission. On the other hand, the spectrum from Ringwald, Naylor and Mukai (1996) is featureless with a red continuum. These authors suggest that it may be heavily reddened or that the detection was negative. 

The closest star in the OGLE database shows variability with a period of 0.1518412(1)~days. The scatter (produced by a flickering) is relatively high ($\sim 0.1$ mag), but we noticed a shallow dimming around the phase 0.5, possibly the secondary eclipse (Figure \ref{fig:eclipsing_post_novae}). Alternatively, the variability might be caused only by ellipsoidal variations of the distorted secondary. In this case, the orbital period would be twice the photometric period, i.e., 0.3036824(3)~days. In the CMD, the star is located blueward of the main sequence.

\subsubsection{V4579 Sgr (Nova Sgr 1986)}

V4579 Sgr was discovered by Robert McNaught on exposures taken between 1986 October 25 and 29 (McNaught 1987). 

The star, as a source of H$\alpha$ emission, was selected as a symbiotic star candidate. Follow-up spectroscopy revealed a blue continuum with strong He II, N III, [Fe VII], and C IV emission lines (Miszalski et al. 2013), typical for He/N-type novae.

We measured a strong period at 0.15356146(4)~days. The light curve (Figure \ref{fig:eclipsing_post_novae}) shows a prominent secondary eclipse.

An $O-C$ analysis shows that the orbital period is increasing at a rate of $(1.1 \pm 0.3) \cdot 10^{-9}$ s s$^{-1}$, which can be used to assess the mass-transfer rate as $\dot{M}=\frac{M_1 M_2}{M_1-M_2}\frac{\dot{P}_{\rm orb}}{3P_{\rm orb}}$ (for conservative mass transfer), where $M_1$ is the mass of primary and $M_2$ that of the secondary. The mass of the red dwarf can be estimated using the assumption that it follows the standard mass-radius relationship and fills its Roche lobe. This leads to the relation between the orbital period of the binary and the secondary mass (see Hellier 2001), which in this case implies $M_2 \approx 0.35\ M_{\odot}$. The white dwarf mass can be estimated roughly from the speed of nova decline. Generally, the more massive the primary, the faster the nova (e.g., Bode \& Evans 2008). Here, we have no information about the speed class of V4579 Sgr, and we assumed $M_1 \approx 0.6\ M_{\odot}$, but we checked that results are similar (within 20\%) in a broader mass range $0.5-0.75\ M_{\odot}$. Under all of these assumptions, we derive $\dot{M} \approx 7\cdot 10^{-7}\ M_{\odot}$ yr$^{-1}$.

The nova is still declining at a rate of $17.4\pm0.5$ mmag yr$^{-1}$.

\subsubsection{V4633 Sgr (Nova Sgr 1998)}
\label{V4633Sgr}

V4633 Sgr was discovered by William Liller in photographs taken on 1998 March 22 (Liller 1998). It had a peak brightness of $V\approx 7.7$ mag. The decline was moderately fast with $t_2\approx 28$~days and $t_3\approx 55$~days (Lynch et al. 2001). Early spectra were dominated by emission lines of the Balmer series, Fe II, N II, and Na I, suggesting that the nova is of the Fe II class (Della Valle et al. 1998). In the infrared regime, prominent lines of H I, C I, O I, and N I were detected (Ashok \& Chandrasekhar 2000). Spectra taken 1.5--2 years after the maximum were typical for this class of object, i.e., with strong coronal lines (Lynch et al. 2001).

X-ray spectra obtained with {\it XMM-Newton} can be interpreted as the emission of multi-temperature, optically thin thermal plasma, possibly heated in shocks (Hernanz \& Sala 2007).

Lipkin \& Leibowitz (2000) noticed two distinct periodicities in the light curve. They interpreted the stable period of 0.1256~days as a signature of orbital frequency. The second period, slightly longer than the orbital one, decreased monotonically from 0.1289~days in 1998 to 0.1277~days in 2003, and eventually disappeared from the light curve (Lipkin et al. 2001; Lipkin \& Leibowitz 2008). The authors favor the interpretation that this system was polar. Before the eruption, due to the strong magnetic field of the primary, the rotation of the white dwarf was synchronized with the orbital revolution. The nova explosion caused a white dwarf expansion, and consequently an increase of the moment of inertia. In analogy to a school example of figure skaters, the white dwarf spin slowed down, leading to the observed asynchronism. After a few years, the white dwarf returned to its initial properties. 

We measured a period of $0.1255667(3)$~days, in agreement with Lipkin \& Leibowitz (2008). We did not find any signatures of other periodicities. The light curve is shown in Figure \ref{fig:eclipsing_post_novae}.

\subsubsection{V4742 Sgr (Nova Sgr 2002b)}

This nova, discovered by William Liller, exploded between 2002 September 11.1 and 15.1 UT (Liller 2002). It declined quickly on timescales of $t_2=12$ days and $t_3=23$ days (Morgan, Ringwald \& Prigge 2003). Early spectra showed emission lines of H I, Fe II, and Na I, typical for Fe II novae (Buil 2002; Duerbeck, Sterken \& Fu 2002). The later spectral and photometric evolution was summarized by Morgan et al. (2003).

OGLE has been monitoring this nova since 2010. Our light curve (Fig. \ref{fig:eclipsing_post_novae}) shows prominent, deep ($\Delta I\sim 3$ mag) eclipses. The orbital period is 0.1336159(1)~days. In principle, it could be twice as long. However, the primary and secondary eclipses would have the same depth, which is unlikely. 

\subsubsection{V5116 Sgr (Nova Sgr 2005b)}

This nova is another discovery by William Liller; the eruption was noticed on 2005 July 4 at $\sim 8$ mag (Liller 2005). The spectrum showed emission lines of H I, He I, C I, N I, Ca II, and O I (Russell et al. 2005). The decline from maximum was very fast with characteristic timescales of $t_2=6.5\pm 1.0$~days, $t_3=20.2\pm 1.9$~days (Dobrotka et al. 2008). Dobrotka et al. (2008) found a photometric periodicity of 0.1238(1)~days and interpreted it as the orbital period. They claim that this nova may have a low mass ratio $q\sim 0.3$, and so the 3:1 resonance may be induced in the accretion disk and give rise to superhumps. They did not find any additional periodicities in the light curve, concluding that the accretion disk was not fully developed to the tidal radius at the time of their observations (acquired in late 2006).

V5116~Sgr regained interest in 2007, two years after the optical maximum, when X-ray observations revealed supersoft spectra with bright emission lines (Nelson et al. 2007; Ness et al. 2007; Sala et al. 2007). The best-fit model consists of an ONe white dwarf with an effective temperature of $\sim 6\cdot 10^5$~K (Sala et al. 2008). The X-ray light curve (showing alternating high and low states with a period of 0.124~days) could be produced by the partial coverage of an asymmetric accretion disk (Sala et al. 2008). 

Models of free-free emission from novae (Hachisu \& Kato 2010) can reproduce the light curve of V5116~Sgr, indicating the mass of an ONeMg white dwarf of $1.07 \pm 0.07 M_{\odot}$. 

We measured an orbital period of 0.1237444(2)~days. We did not find any additional periodicities in the light curve, which rules out the presence of superhumps in this system. The out-of-eclipse brightness is still declining at a rate of $99.8 \pm 8.4$ mmag yr$^{-1}$. The light curve is shown in Figure \ref{fig:eclipsing_post_novae}.

\subsubsection{V5627 Sgr}

This star is known in the literature as ''MACHO peculiar variable'' (because it was found in the data from the MACHO project; T. Kato 2000, vsnet-chat 3454). It showed a suspicious linear decline from $V=15.1$ to $V=16.9$ mag, resembling a late decline after nova eruption. The Downes et al. (2001) catalog has a short note that this object was in outburst on a POSS II plate (taken on 1991 September 6), however, we did not find any references.

A high-speed photometry showed a strong modulation with a period of 0.1170(5)~days (Woudt et al. 2005). However, the light curve was not stable, displaying appearing and disappearing secondary eclipses. The spectrum is dominated by strong He II, N III, [Fe VII], and C IV emission lines (Miszalski et al. 2013) resembling those of He/N-type novae.

V5627 Sgr has been continuously monitored by OGLE since 2001. It is still declining at a constant rate of $30.5\pm 1.6$ mmag yr$^{-1}$. We measured an orbital period of 0.117161(2)~days in agreement with the value from Woudt et al. (2005). The light curve (see Figure \ref{fig:eclipsing_post_novae}) shows primary eclipses with an amplitude of $~\sim 0.15$ mag and probable shallow secondary eclipses.

\subsubsection{V825 Sco (Nova Sco 1964)}

V825 Sco is a poorly studied nova. It was discovered by Antoni Przybylski on 1964 May 19, during a decline phase from maximum (Przybylski 1964). Wilde (1965) estimated that it peaked at $V\sim 8$ mag in December 1963. Available spectra (taken 28 and 131 days after the discovery) show prominent lines of [O III] and [Ne III] (Wilde 1965).

The remnant shows deep eclipses ($\Delta I=1.5$ mag) with a period of 0.19165876(3) days. The secondary eclipses are shallow (Figure \ref{fig:eclipsing_post_novae}). In the CMD, the star is located blueward of the main sequence.

\section{On the Galactic bulge nova rate}
\label{sec:rate}

Direct measurements of the nova rate in the Milky Way are subject to many uncertainties and biases. First, due to observational bias, many eruptions may be missed, although they are potentially observable by a survey. There are many of reasons for that. A nova eruption might have occurred during a conjunction with the Sun (or the Full Moon) or telescope maintenance. The explosion might have taken place in a sky region that was not monitored. Alternatively, observers might have recorded the outburst but not recognized and alerted it. It is particularly hard to quantify properly this bias, especially when observations from different sources are used. Usually, one assumes that all very bright novae were recorded and corrects their number for the search efficiency.

The second caveat is to estimate correctly what fraction of all Galactic novae {\it could} be observed. This number depends on the distribution of pre-novae in the Milky Way and on the distribution of interstellar matter, often patchy and non-uniform. The extinction might be especially troublesome in visual bands, and so observations in the near-infrared and infrared regimes are needed to lower its influence.

The first estimates of the nova rate yielded very high values: $\sim 100$~yr$^{-1}$ (Allen 1954), $\sim 50$~yr$^{-1}$ (Kopylov 1955), $\sim 260$~yr$^{-1}$ (Sharov 1972), and $\sim 73$~yr$^{-1}$ (Liller \& Mayer 1987), which disagreed with observations of neighboring galaxies. Recent calculations lowered this value. Hatano et al. (1997) used Monte Carlo simulations to reanalyze Liller \& Mayer (1987) data and found a nova rate of $41 \pm 20$ \peryr with the majority of eruptions originating in the disk population. Shafter (1997) measured a rate of $35 \pm 11$ \peryr. However, both authors assumed that the distribution of novae is axisymmetric about the Galactic center. Shafter (1997) noted that the presence of a Galactic bar may alter this results significantly.

Another, indirect approach is based on observations of extragalactic novae. The nova rate seems to be correlated with the brightness of the parent galaxy. The first results constrained the Galactic nova rate in a broad range 11--50 \peryr ~  (Ciardullo et al. 1990; Della Valle \& Livio 1994). Darnley et al. (2006) estimated in this way the Galactic rate of $34^{+15}_{-12}$ \peryr ~ with a bulge rate of $14^{+6}_{-5}$ \peryr. 

In this section, we would like to revisit this problem. First, we have a large statistical sample of novae observed over several years in a large sky area. Second, we have used a uniform set of criteria to (successfully) identify novae in our data set which allows us to minimize observational biases. Finally, we will adopt a modern model of the Galactic bulge and interstellar extinction.

\subsection{Observed Galactic bulge nova rate}

Because of the much larger spatial coverage of OGLE-III and -IV compared to previous phases of the project (Section \ref{sec:data}), we will limit our considerations to novae that erupted between 2001 and 2013. OGLE-III fields in the Galactic bulge covered 92 sq. degrees, while OGLE-IV observations analyzed here extend to a larger region of 140 sq. degrees. They cover practically the entire inner Galactic bulge area, in the $-10^{\circ}<l<10^{\circ}$ and $-7^{\circ}<b<5^{\circ}$ region. 

Eruptions of novae have high amplitudes which make them easy to detect. We believe that our search procedures have enabled us to discover practically all novae brighter than $I_{\rm min}=17$ mag in maximum. Fainter objects could have been mistaken for other erupting variables, and for this reason we omit them in our calculations.\footnote{We note that the survey detection limit is $I\sim21$ mag. Hence, the star peaking at $I\approx 18$ mag might be an ''ordinary,'' nearby dwarf nova of WZ Sge-type or a distant and/or heavily reddened CN. This may lead to confusion, especially when the cadence is poor. This is the case in fields along the Galactic equator where novae are heavily reddened.
}

We also neglect four novae (V2574 Oph, V2576 Oph, V2615 Oph, and V5580 Sgr), for which only late declines were recorded; all of these were known from the literature. We end up with 30 objects, which gives an observed nova rate of $2.3 \pm 0.5$ \peryr (the Poissonian error was assumed). We note that almost 1/3 of these are OGLE-based discoveries: they were relatively bright, but eruptions occurred around the seasonal gap. Restricting ourselves to the first four years of the OGLE-IV yields a much higher observed rate of $4.8 \pm 1.1$ \peryr.

\subsection{Discovery probability}

Only a (small?) fraction of the bulge novae could be detected by the OGLE survey. In order to find a value of this efficiency factor, we have to make essentially {\it one} assumption about the nova distribution, namely, that it follows the luminosity density of the Galactic bulge. This assumption is fulfilled for M31 and other neighboring galaxies (see Bode \& Evans 2008) and was used in previous estimates of the nova rate. 

The outline is as follows. We will produce a mock catalog of novae from a realistic distribution and pick up those objects falling inside the OGLE coverage and brightness range. For each of these we will calculate the probability of detection based on the time vectors of observations in a given field, which will will finally lead to the value of the efficiency factor.

In the first approximation, the Galactic bulge can be described as a triaxial ellipsoid, slightly inclined to the line of sight. However, the precise description of its shape and composition is a matter of active research. For example, the latest studies revealed the presence of an X-shaped (peanut-shaped) structure (e.g., McWilliam \& Zoccali 2010; Nataf et al. 2010). There is an evidence that the bar inclination increases in the inner bulge (Pietrukowicz et al. 2015).

We used the Galaxia code (Sharma et al. 2011) to generate a catalog of stars from the Galactic bulge. In that model, the stellar density follows $\rho(x,y,z) \propto \exp (-0.5 r^2_s)$, where $r_s^4 = ((x/x_0)^2+(y/y_0)^2)^2+(z/z_0)^4$ (with $x_0 = 1.59$~kpc and $y_0=z_0=0.424$~kpc) and this distribution is truncated at 2.54~kpc from the Galactic center. Sharma et al. also used a small inclination angle of $11^{\circ}$ (which is the angle between the major axis of the bar and the line connecting the Sun and the Galactic center). The parameters of this distribution were derived from the near-infrared star counts from the DENIS survey and were used in the classic Besan\c{c}on model (Robin et al. 2003). 

The extinction was calculated assuming that the distribution of the interstellar material may be approximated by the double exponential function $\rho_{\rm ism} \propto \exp (-(R-R_0)/h_R) \exp (-|z|/h_z)$, where $h_R = 4.2$~kpc and $h_z = 88$~pc are characteristic scale-lengths, and $R_0 = 8.0$~kpc is the distance to the Galactic center (Sharma et al. 2011). The normalization constant was chosen to produce the value of $E(J-K_s)$ toward a given direction from the VVV extinction maps (Gonzalez et al. 2012). These maps cover practically all of the bulge ($-10^{\circ}\leq l \leq 10.2^{\circ}$, $-10^{\circ}\leq b \leq 5^{\circ}$). If the star fell outside the VVV coverage, we used the Schlegel et al. (1998) maps instead ($\sim 15$\% of cases). We adopted the Cardelli extinction law (Cardelli et al. 1989).

We also assumed that the Galactic bulge novae have a Gaussian luminosity function with a mean absolute brightness of $M_I=-7$ mag and a standard deviation of 1 mag. This is in agreement with the apparent brightness of extragalactic novae. Nonetheless, we checked that other reasonable luminosity functions provide consistent results (see discussion).

Finally, for each star, we calculated the probability of detection in the OGLE survey. We used the time vectors of observations to create artificial light curves: we assumed that the nova is detectable for 20 days after the eruption (below, we discuss the impact of this parameter on final results). We estimated this value based on light curves of novae from our sample. We required at least three observations (three nights) in outburst to ''detect'' it (this condition was used in the searching procedure). Assuming that eruptions are distributed uniformly in time, we calculated ''the detection efficiency,'' i.e., the probability of detection in our survey. We plot this function in Figure \ref{fig:discovery_efficiency}. For observations from 2001-2013, the highest efficiency is 60-70\%. In the OGLE-IV phase (since 2010), our observing capabilities have increased, resulting in a higher detection efficiency (up to 80\%) in a much larger bulge area. Since these sky regions are not available for observations for three months annually, this is the maximal value that could be obtained.

Combining all of the ingredients together, we calculate the efficiency factor. During 2010--2013, we detected 36.1\% of all bulge novae, while during 2001--2013 this fraction was 16.7\%. Finally, we obtain the true Galactic bulge nova rates of $13.2 \pm 3.1$~\peryr~ (2010--2013) and $13.8 \pm 2.6$~\peryr~ (2001--2013). Both values are consistent at the $1\sigma$ level. This result also confirms estimations by Darnley et al. (2006), who assessed the Galactic bulge nova rate of $14^{+6}_{-5}$ \peryr~based on observations of M31 novae.

\subsection{Discussion}

In the previous section, we have made a number of assumptions about the nova distribution. Now, we would like to discuss their applicability, limitations, and impact on the final results.

First, we compared observed and model distributions of novae in the sky using the two-dimensional generalization of the Kolmogorov--Smirnov test (Fasano \& Franceschini 1987). We obtained the $p$-value in the range 0.13--0.25, depending on a model sample, meaning that the null hypothesis (that both distributions are the same) cannot be rejected at a significance level of $\alpha = 0.13$ or smaller. In other words, there is no evidence that the observed and model distributions are different. 

We assumed that all 30 objects that were used here to compute the observed nova rate belong to the bulge population. There are still discussions of whether or not nova properties depend on the underlying population. Generally, it is believed that Galactic disk novae are brighter and faster than those from the Galactic bulge. With data from the bulge only, we are unable to separate both populations. This will be possible in the near future, when observations from the OGLE-IV Galaxy Variability Survey will be available. However, we estimate that the contamination from the disk population is practically negligible. We observe in a narrow longitude range $-10^{\circ}<l<10^{\circ}$, thus, assuming that we cannot observe objects behind the bulge, the contamination from the disk is $5-10$ \% at most. This automatically lowers the nova rate by $5-10$ \%, which is still within the $1\sigma$ errorbar.

The model of the Galactic bulge, which is used in the Galaxia code (and other codes, like Besan\c{c}on model), uses a small value of inclination angle of $i=11^{\circ}$, between the major axis of the bar and the line toward the Galactic Center. The latest observations indicate that this angle is larger, $i\approx 20^{\circ}$ (e.g., Pietrukowicz et al. 2015). In order to quantify this effect, we repeated calculations from the previous subsection. We used a larger inclination angle $i= 20^{\circ}$ but kept the other parameters fixed. This may not be the case, since the default bulge parameters in the code were fit to the observations (star counts in the infrared range) and their values might be affected by correlations. 

One would expect that in the model with larger inclination, stars would be spread over a larger sky area, and hence the derived detection efficiency would be overestimated. Indeed, the detection probability would be 14.5\% ($\sim 10$\% smaller than in the previous subsection). The final nova rate would be slightly ($\sim 10$\%) higher. 

We note that both effects (contamination from the disk population and higher inclination of the bar) act in opposite directions, and hence nearly cancel each other out.

The impact of extinction on the final result is surprisingly small. Only $\sim1$\% of potentially observable objects were rejected because of high reddening. This is caused by two reasons: (1) we use photometry in the $I$ band, where extinction is smaller than in the visual bands; and (2) regions of high extinction were observed at poor cadence, and so the chances of detection are small whatsoever. If there was no extinction at all, then the nova rate would be $\sim 1$\% higher.

Similarly, the effect of the novae luminosity function will be small. We tested other, broader or narrower Gaussian distributions and Delta-like functions centered on $-8$, $-7$, and $-6$ mag. In all cases, the derived nova rate differs by 2\%--3\% from that calculated in the previous section. 

The choice of a threshold peak brightness of novae has practically no impact on our final results. For $I_{\rm min}$ in a range 15-19 mag, the detection probability would change by less than 1\%. The saturation limit of the survey is $\sim 11-12$ mag (Udalski et al. 2015), and so the brightest novae may be missed. However, we found that the contribution of novae brighter than 6 mag in peak to the overall bulge rate is less than $\sim2$ \%.

We assumed that the nova is detectable for 20 days after the eruption. This is consistent with the fact that the Galactic bulge novae are generally slower than those from the disk. However, the choice of this parameter does not significantly affect the final results. For 15~days, we obtained a discovery probability of 14.8\%, for 30~days 18.1\%. This changes the final nova rate by $\pm 10$\%, which is still within the $1\sigma$ errorbar. Even for the fastest novae, like V745 Sco, the detection probability does not vary significantly. If we assume that the nova is detectable for only 10 days after the eruption, then we obtain an upper limit on the bulge nova rate of 19.5 \peryr, which is marginally consistent (at $2.2\sigma$) with the derived nova rate. On the other hand, if we could discover all of the novae in our fields, then the bulge nova rate would be 7.7 \peryr. This lower limit is consistent (at $2.3\sigma$) with the value which we measured.

\begin{figure*}
\centering
\includegraphics[width=0.75\textwidth]{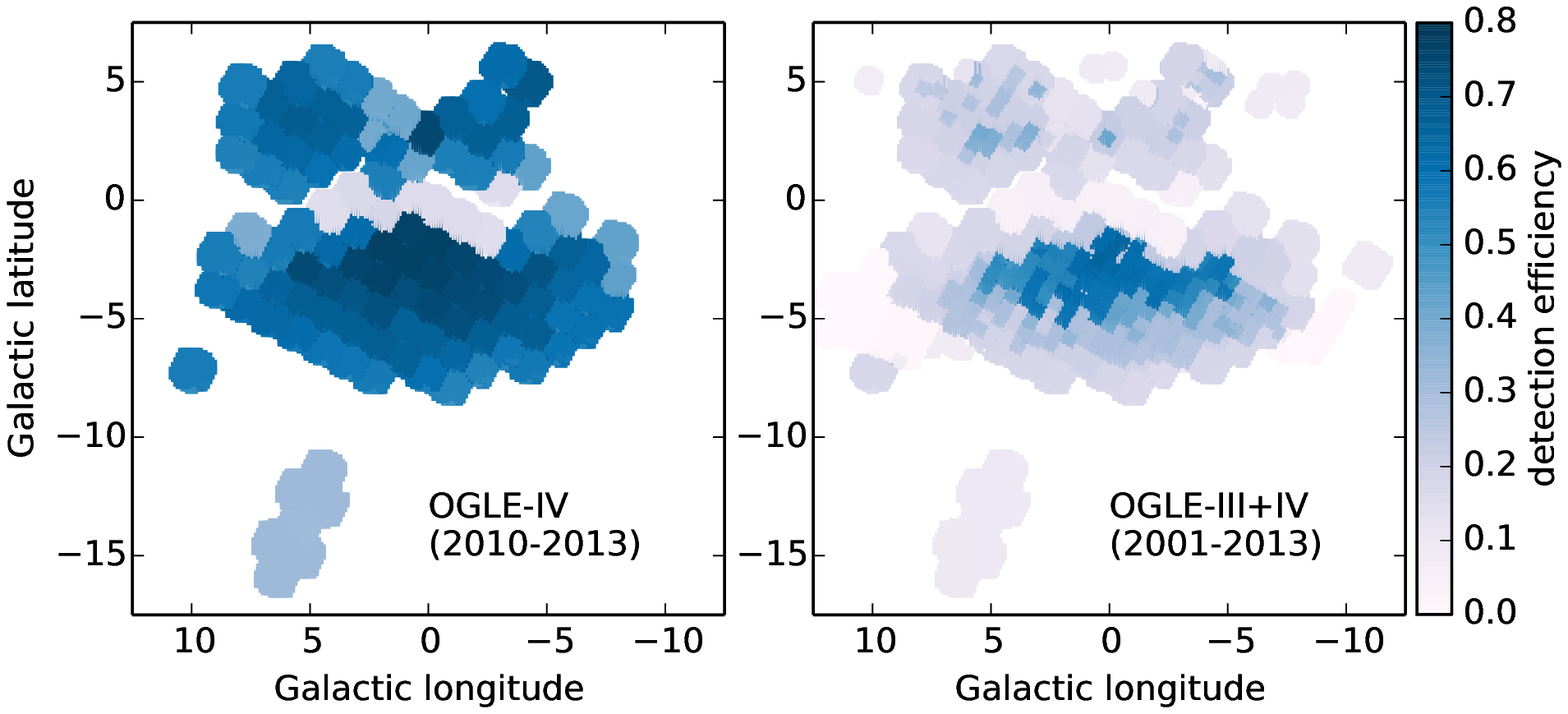}
\caption{Nova detection efficiency in OGLE-monitored fields in 2010-2013 (left panel; OGLE-IV only) and 2001-2013 (right panel; OGLE-III+IV). We assumed that the nova can be detected for 20 days after the eruption.}
\label{fig:discovery_efficiency}
\end{figure*}

\section{A number of recurrent novae}
\label{section:howmanyRN}

Recently, Darnley et al. (2014) and Tang et al. (2014) announced the discovery of the remarkable nova M31 2008-12a. They recorded six eruptions of this nova in six consecutive years. Such a short recurrence timescale may be observed only if the white dwarf is extremely massive, close to the Chandrasekhar limit ($M>1.3\ M_{\odot}$), and the mass-transfer rate is very high ($\dot{M}>1.5\cdot 10^{-7} M_{\odot}\rm{yr}^{-1}$; see Kato et al. 2014).

How many similar novae with a very short recurrence timescale are in the Milky Way? Since we have not detected any, we are only able to set upper limits on their number. We repeated our calculations from Section \ref{sec:rate}. For a given recurrence period $\tau_{\rm rec}$, we measured the probability of detection of {\it at least} two eruptions between 2001 and 2013. We also assumed that the eruption is observable for 10 days, since the more frequent eruptions, the faster the nova. For $\tau_{\rm rec}=300$~days the detection probability is 40\%, for $\tau_{\rm rec}=500$~days it is 28\%, for $\tau_{\rm rec}=1000$~days it is 14\%. The shorter the recurrence timescale, the higher the probability of detection. Stronger constraints may be obtained if we assume that the nova outburst is detectable for five days only (see Figure 2 of Darnley et al. 2014). In such a case, for $\tau_{\rm rec}=1000$~days, the detection probability drops to 6.8\%. In other words, the number of novae (in the Galactic bulge) with a recurrence timescale shorter than three years is a dozen at most. 

We cannot exclude that some of them may end their life as SNe Ia. However, their number is too small to produce the observed SN Ia rate ($(7.2 \pm 2.3)\cdot 10^{-3}\ \rm{yr}^{-1}$; Maoz et al. 2014). For example, for a white dwarf mass of $1.3\ M_{\odot}$ and a mass-transfer rate of $3\cdot 10^{-7}\ M_{\odot}\rm{yr}^{-1}$ (nova eruptions  would be observed every year), the time needed to reach the Chandrasekhar mass is at least $3\cdot 10^6$ years (assuming that only 10\% of mass is accumulated). Thus, the death rate of such novae is at most $3\cdot 10^{-6}\ \rm{yr}^{-1}$, meaning that their contribution to the overall SNe Ia rate is negligible.

We would like to emphasize that this does not mean that RNe are not SN Ia progenitors. We explored only a small fraction of the parameter space, namely, systems with extremely massive white dwarfs and very high mass-transfer rate. Such stars are expected to erupt very frequently, at least once per a few years. The contribution of recurrent novae with inter-eruption timescales 3--100~years may still be significant (Pagnotta \& Schaefer 2014). For example, for $\tau_{\rm rec}=5$~years the detection probability is 7.0\%, for $\tau_{\rm rec}=10$~years it is 0.9\%.

We found that the population of frequently erupting RNe in the Galactic bulge is negligible, a few dozens at most (if any). This means that an evolutionary scenario that leads to similar systems is very unlikely. Thus, the detection of a one year recurrent nova (like the case of M31 2008-12a) would indicate the discovery of a very unique object.

\section{Summary}

In this paper, we have presented a large collection of CNe from the Galactic bulge, observed over 23 years during the course of the OGLE sky survey. We believe that the data described here can serve many purposes. First, they allow studies of individual rare phenomena. We presented a detailed photometric evolution of rebrightenings after the eruption of V5585 Sgr (Nova Sgr 2010). We identified two rare novae, V5587 Sgr (Nova Sgr 2011) and OGLE-2006-NOVA-01, which showed cusp-shaped secondary maxima. Second, we provide information about the overall population of bulge novae. We measured orbital periods for 18 post-novae and novae, noticeably increasing the number of such systems. We have used a uniform set of criteria to identify novae in our data set, which allowed us to measure a Galactic bulge nova rate of $13.8 \pm 2.6$ \peryr. We also showed that RNe with short inter-eruption timescales (less than three years) are extremely rare.

We plan to extend our analyses for CNe to the Magellanic Clouds and the Galactic Disk.

\section*{Acknowledgments}

The OGLE project has received funding from the National Science Centre, Poland, grant MAESTRO 2014/14/A/ST9/00121 to AU.
P.M. is supported by the ''Diamond Grant'' No. DI2013/014743 funded by the Polish Ministry of Science and Higher Education.
This work has been supported by the Polish National Science Centre grant No. DEC-2011/03/B/ST9/02573. This research has made use of the International Variable Star Index (VSX) database, operated at AAVSO, Cambridge, Massachusetts, USA, and the SIMBAD database, operated at CDS, Strasbourg, France.

\begin{table*}
\centering
\caption{List of Nova Eruptions Recorded by the OGLE Survey. }
\begin{tabular}{lr@{$^{\rm{h}}$}r@{$^{\rm{m}}$}r@{\fs}rr@{$^{\circ}$}r@{$'$}r@{\farcs}rllr}
\hline
\multicolumn{1}{c}{Designation} & \multicolumn{4}{c}{R.A.$_{\rm J2000.0}$} & \multicolumn{4}{c}{Decl.$_{\rm J2000.0}$} & Field & Star ID & \multicolumn{1}{c}{Date} \\ \hline
\object{Nova WISE}         & 18 & 18 & 34 & 02 & -28 & 49 & 19 & 8 & BLG532.30 & 84389  & 2010-03-24 \\ 
\object{OGLE-1997-NOVA-01}* & 17 & 50 & 36 & 08 & -30 & 01 & 46 & 6 & BUL\_SC5  & 244353 & 1997-04-13 \\ 
\object{OGLE-2006-NOVA-01}* & 18 & 00 & 49 & 71 & -30 & 48 & 58 & 8 & BLG183.1  & 143458 & 2006-02-10 \\ 
\object{OGLE-2008-NOVA-01}* & 17 & 56 & 10 & 35 & -30 & 04 & 36 & 5 & BLG103.6  & 53988  & 2008-02-10 \\ 
\object{OGLE-2010-NOVA-01}* & 17 & 54 & 34 & 76 & -23 & 32 & 21 & 5 & BLG643.14 & 42081  & 2010-06-26 \\ 
\object{OGLE-2010-NOVA-02}* & 17 & 36 & 59 & 70 & -29 & 08 & 14 & 9 & BLG653.02 & 34242  & 2010-10-03 \\ 
\object{OGLE-2010-NOVA-03}* & 18 & 05 & 25 & 36 & -31 & 54 & 05 & 1 & BLG515.26 & 74708  & 2010-03-05 \\ 
\object{OGLE-2011-NOVA-01}** & 17 & 32 & 23 & 10 & -29 & 48 & 38 & 5 & BLG613.01 & 1N     & 2011-11-16 \\ 
\object{OGLE-2011-NOVA-02}* & 17 & 36 & 59 & 57 & -29 & 51 & 56 & 2 & BLG654.10 & 35091  & 2011-02-10 \\ 
\object{OGLE-2012-NOVA-01}** & 17 & 56 & 49 & 39 & -27 & 13 & 28 & 2 & BLG645.05 & 124N   & 2012-05-02 \\ 
\object{OGLE-2013-NOVA-04}* & 18 & 01 & 22 & 20 & -25 & 18 & 39 & 0 & BLG664.06 & 45N    & 2013-03-31 \\ 
\object{OGLE-2014-NOVA-01}** & 17 & 55 & 20 & 42 & -23 & 23 & 54 & 9 & BLG643.22 & 8N     & 2014-02-16 \\ 
\object{V2574 Oph} (Nova Oph 2004)  & 17 & 38 & 45 & 51 & -23 & 28 & 18 & 5 & BLG626.31 & 130884 & 2004-04-14 \\ 
\object{V2576 Oph} (Nova Oph 2006b) & 17 & 15 & 32 & 99 & -29 & 09 & 40 & 1 & BLG617.30 & 22933  & 2006-04-06 \\
\object{V2615 Oph} (Nova Oph 2007)  & 17 & 42 & 44 & 00 & -23 & 40 & 35 & 2 & BLG632.16 & 123083 & 2007-03-22 \\ 
\object{V2670 Oph} (Nova Oph 2008)  & 17 & 39 & 50 & 93 & -23 & 50 & 00 & 9 & BLG626.22 & 59141  & 2008-05-25 \\ 
\object{V2673 Oph} (Nova Oph 2010)  & 17 & 39 & 40 & 98 & -21 & 39 & 47 & 7 & BLG624.13 & 51204  & 2010-01-15 \\ 
\object{V2674 Oph} (Nova Oph 2010b) & 17 & 26 & 32 & 11 & -28 & 49 & 38 & 7 & BLG614.09 & 78515  & 2010-02-18 \\ 
\object{V2677 Oph} (Nova Oph 2012b) & 17 & 39 & 57 & 01 & -24 & 47 & 07 & 1 & BLG714.30 & 16752  & 2012-05-19 \\ 
\object{V4444 Sgr} (Nova Sgr 1999)  & 18 & 07 & 36 & 21 & -27 & 20 & 13 & 2 & BUL\_SC19 & 62367  & 1999-04-25 \\ 
\object{V4740 Sgr} (Nova Sgr 2001c) & 18 & 11 & 45 & 89 & -30 & 30 & 50 & 0 & BLG187.3  & 132611 & 2001-09-05 \\ 
\object{V4741 Sgr} (Nova Sgr 2002)  & 17 & 59 & 59 & 67 & -30 & 53 & 20 & 8 & BLG183.8  & 59087  & 2002-04-15 \\ 
\object{V4741 Sgr} (Nova Sgr 2002)  & 17 & 59 & 59 & 63 & -30 & 53 & 20 & 8 & BLG514.25 & 119898 & 2002-04-15 \\ 
\object{V5113 Sgr} (Nova Sgr 2003b) & 18 & 10 & 10 & 35 & -27 & 45 & 33 & 0 & BLG227.2  & 146825 & 2003-09-17 \\ 
\object{V5579 Sgr} (Nova Sgr 2008)  & 18 & 05 & 58 & 89 & -27 & 13 & 56 & 1 & BLG234.2  & 39156  & 2008-04-18 \\ 
\object{V5579 Sgr} (Nova Sgr 2008)  & 18 & 05 & 58 & 89 & -27 & 13 & 56 & 1 & BLG511.17 & 60064  & 2008-04-18 \\ 
\object{V5580 Sgr} (Nova Sgr 2008b) & 18 & 22 & 01 & 48 & -28 & 02 & 39 & 6 & BLG531.08 & 124387 & 2008-11-29 \\ 
\object{V5582 Sgr} (Nova Sgr 2009b) & 17 & 45 & 05 & 41 & -20 & 03 & 21 & 6 & BLG629.12 & 58287  & 2009-02-23 \\ 
\object{V5583 Sgr} (Nova Sgr 2009c) & 18 & 07 & 07 & 70 & -33 & 46 & 34 & 6 & BLG597.32 & 7632   & 2009-08-06 \\ 
\object{V5585 Sgr} (Nova Sgr 2010)  & 18 & 07 & 26 & 95 & -29 & 00 & 43 & 7 & BLG520.23 & 28771  & 2010-01-20 \\ 
\object{V5586 Sgr} (Nova Sgr 2010b) & 17 & 53 & 03 & 05 & -28 & 12 & 18 & 8 & BLG500.27 & --     & 2010-04-23 \\ 
\object{V5587 Sgr} (Nova Sgr 2011)  & 17 & 47 & 46 & 22 & -23 & 35 & 13 & 5 & BLG632.09 & 56209  & 2011-03-07 \\ 
\object{V5588 Sgr} (Nova Sgr 2011b) & 18 & 10 & 21 & 35 & -23 & 05 & 30 & 1 & BLG717.09 & 77655  & 2011-03-27 \\ 
\object{V5589 Sgr} (Nova Sgr 2012)  & 17 & 45 & 28 & 03 & -23 & 05 & 22 & 7 & BLG632.29 & 128028 & 2012-04-21 \\ 
\object{V5591 Sgr} (Nova Sgr 2012c) & 17 & 52 & 25 & 80 & -21 & 26 & 21 & 7 & BLG637.18 & 115556 & 2012-06-26 \\ 
\object{V5592 Sgr} (Nova Sgr 2012d) & 18 & 20 & 27 & 28 & -27 & 44 & 26 & 8 & BLG531.20 & 66951  & 2012-07-07 \\ 
\object{V745 Sco}  (Nova Sco 1937)  & 17 & 55 & 22 & 22 & -33 & 14 & 58 & 6 & BLG508.07 & 67742  & 2014-02-06 \\ 
\object{V1141 Sco} (Nova Sco 1997)  & 17 & 54 & 11 & 22 & -30 & 02 & 52 & 6 & BUL\_SC4  & 31705  & 1997-06-05 \\ 
\object{V1178 Sco} (Nova Sco 2001)  & 17 & 57 & 06 & 66 & -32 & 23 & 05 & 4 & BLG157.6  & 75141  & 2001-05-13 \\ 
\object{V1188 Sco} (Nova Sco 2005)  & 17 & 44 & 21 & 63 & -34 & 16 & 35 & 2 & BLG129.2  & 126282 & 2005-07-27 \\ 
\object{V1533 Sco} (Nova Sco 2013)  & 17 & 33 & 59 & 47 & -36 & 06 & 19 & 4 & BLG672.28 & 23585  & 2013-02-02 \\ 
\hline
\end{tabular}\\
\begin{flushleft}
Notes:\\
For each object, we list designation, equatorial coordinates (for the epoch J2000.0), the OGLE identifier, and the date of the eruption. \\
* Novae discovered in the OGLE archival data.\\
** Novae discovered by the OGLE survey in a real-time.
\end{flushleft}
\label{tab:novae}
\end{table*}

\begin{table*}
\centering
\caption{Properties of Novae Recorded by the OGLE Survey. }
\begin{tabular}{lcrrrrr}
\hline
\multicolumn{3}{c}{} & \multicolumn{2}{c}{Pre-nova} & \multicolumn{2}{c}{Post-nova} \\
\multicolumn{1}{c}{Designation} & \multicolumn{1}{c}{Type} & \multicolumn{1}{c}{$P_{\rm orb}$} & \multicolumn{1}{c}{$I_{\rm pre}$} & \multicolumn{1}{c}{$(V-I)_{\rm pre}$} & \multicolumn{1}{c}{$I_{\rm post}$} & \multicolumn{1}{c}{$(V-I)_{\rm post}$} \\ \hline
Nova WISE         & S &            & 19.74 & --   & 18.80 & -1.12 \\
OGLE-1997-NOVA-01 & S &            & --    & --   & 19.48 &  --   \\
OGLE-2006-NOVA-01 & C &            & $>21$ &      & $>21$ &  --   \\
OGLE-2008-NOVA-01 & S &            & 19.67 & --   & 19.98 & --    \\
OGLE-2010-NOVA-01 & S &            & 21.44 &  --  & 21.18 & --    \\
OGLE-2010-NOVA-02 & O &            & $>21$ &   -- &  17.80$\ddagger$ & --   \\
OGLE-2010-NOVA-03 & S & 1.2664(1)   & 18.80 & --   & 18.21 & 0.83  \\
OGLE-2011-NOVA-01 & D &            & $>20.15$ &  --  & 19.65$\ddagger$ & --    \\
OGLE-2011-NOVA-02 & O &            & 21.28 & 1.99 & 21.42 & 2.29  \\
OGLE-2012-NOVA-01 & P &            & $>20.21$ &  --  & 19.56$\ddagger$ & --    \\
OGLE-2013-NOVA-04 & O &            & $>20.25$ &  --  & 18.11$\ddagger$ & --    \\
OGLE-2014-NOVA-01 & S &            & $>20.18$ &  --  & 19.48$\ddagger$ & --    \\
V2574 Oph         & -- &           &  --   &  --  &   18.20$\ddagger$     &  --  \\
V2576 Oph         & -- &            & 19.50 & --   & 18.99 & 0.99  \\
V2615 Oph         & -- & 0.272339(1) & 18.74 & --   & 18.40 & 1.27  \\
V2670 Oph         & S &            &  --   & --   &  --   & --    \\
V2673 Oph         & S &            &   --  &  --  & 20.93 & --    \\
V2674 Oph         & S & 1.30207(6)  &  --   &  --  & 19.47 & 2.06  \\
V2677 Oph         & S & 0.344295(8) & 19.81 &  --  &  19.49 & --    \\
V4444 Sgr         & S &            &  --   & --   & 19.35 & 1.73  \\
V4740 Sgr         & S &            & --    & --   & 19.12 & 0.47  \\
V4741 Sgr         & S &            & 18.70 & --   & 18.66 & 1.89  \\
V5113 Sgr         & S & 0.150015(1) & 20.60 & --   & 20.02 & 0.71  \\
V5579 Sgr         & D &            & 20.31 & --   & 17.00$\ddagger$ &   --  \\
V5580 Sgr         & -- &             &  --   & --   & 17.49 & 1.01  \\
V5582 Sgr         & P & 0.156604(1) & 19.43 & --   & 17.95$\ddagger$ & 0.17  \\
V5583 Sgr         & S &            & 20.08 & 0.65 & 19.72 & 0.59  \\
V5585 Sgr         & J & 0.137526(2) & $>21$ & -- &       17.77 & 0.54 \\
V5586 Sgr         & S &            & 19.92 & --   & 19.78 &   --  \\
V5587 Sgr         & C &            & $>19.5$ & 2.11    & 18.46 & --    \\
V5588 Sgr         & J & 0.214321(8) &  --   &  --  & 17.74 &  --   \\
V5589 Sgr         & S & 1.59230(5)  & 16.95 & 2.54 & 17.00 & 3.05  \\
V5591 Sgr         & S &            & 18.56 &  --  & 17.80 &  --   \\
V5592 Sgr         & D &            & 19.32 & 1.22 & 16.80$\ddagger$      & --    \\
V745 Sco          & S &            & 13.78 & 4.68 & 13.78 &  --   \\
V1141 Sco         & S &            & 18.79 & --   & 18.50 &  --   \\
V1178 Sco         & S &            & --    & --   & 19.54 & 0.39  \\
V1188 Sco         & S &            & 18.92 & --   & -- & --    \\
V1533 Sco         & S &            & 20.03 &  --  & 19.70$\ddagger$    & --    \\

\hline
\end{tabular}
\begin{flushleft}
Notes: \\
For each object, we list the light curve type$\dagger$, orbital period (if measured), mean $I$-band brightness and $(V-I)$ color of pre- and post-nova.\\
$\dagger$ Light-curve type according to the Strope et al. (2010) classification scheme: S - smooth decline, P - plateaus, D - dust dips, C - cusp-shaped secondary maxima, O - oscillations, F - flat-topped, J - jitters and flares. \\$\ddagger$ During decline.
\end{flushleft}
\label{tab:novae_properties}
\end{table*}

\begin{table*}
\centering
\caption{Likely Post-novae: Stars Showing Eclipse-like Variability.}
\begin{tabular}{llllllllllll}
\hline
GCVS ID & Nova & RA$_{\rm J2000.0}$                      & DEC$_{\rm J2000.0}$            & d [$''$] & Field & Star ID & $I$ & $V$ & $V-I$ & Donor type & $P_{\rm orb}$ [d]\\
\hline
\object{V630 Sgr} & Sgr 1936c & 18$^{\rm{h}}$08$^{\rm{m}}$48\fs 39 & -34$^{\circ}$20$'$21\farcs 7 & 1.63 & BLG597.13 & 9738  & 17.228 & 17.736 & 0.508 & MS & 0.1179304(2)\\
\object{V726 Sgr} & Sgr 1936b & 18$^{\rm{h}}$19$^{\rm{m}}$33\fs 67 & -26$^{\circ}$53$'$19\farcs 7 & 0.31 & BLG530.12 & 47046  & 18.254 & 19.395 & 1.141 & MS & 0.822812(3)\\
\object{V999 Sgr} & Sgr 1910  & 18$^{\rm{h}}$00$^{\rm{m}}$05\fs 55 & -27$^{\circ}$33$'$13\farcs 9 & 1.42 & BLG511.16 & 106376 & 16.385 & 17.035 & 0.650 & MS & 0.1518412(1)\\
\object{V1174 Sgr} & Sgr 1952b & 18$^{\rm{h}}$01$^{\rm{m}}$37\fs 03 & -28$^{\circ}$44$'$19\farcs 0 & 7.00 & BLG512.14 & 146226 & 18.502 & -- & -- & -- & 0.3090452(2)\\
\object{V4579 Sgr} & Sgr 1986  & 18$^{\rm{h}}$03$^{\rm{m}}$37\fs 91 & -28$^{\circ}$00$'$07\farcs 8 & 1.13 & BLG512.28 & 169115 & 16.354 & 17.382 & 1.029 & MS & 0.15356146(4) \\
\object{V4579 Sgr} & Sgr 1986  & 18$^{\rm{h}}$03$^{\rm{m}}$37\fs 91 & -28$^{\circ}$00$'$07\farcs 8 & 1.15 & BLG511.03 & 92334  & 16.381 & --     & --    & -- & 0.15356146(4) \\
\object{V4633 Sgr} & Sgr 1998  & 18$^{\rm{h}}$21$^{\rm{m}}$40\fs 49 & -27$^{\circ}$31$'$37\farcs 3 & 0.06 & BLG531.26 & 22870  & 19.366 & 19.976 & 0.610 & MS? & 0.1255667(3) \\
\object{V4742 Sgr} & Sgr 2002b & 18$^{\rm{h}}$02$^{\rm{m}}$21\fs 86 & -25$^{\circ}$20$'$32\farcs 2 & 0.05 & BLG664.05 & 13499  & 19.324 & 21.037 & 1.714 & MS & 0.1336159(1) \\
\object{V5116 Sgr} & Sgr 2005b & 18$^{\rm{h}}$17$^{\rm{m}}$50\fs 77 & -30$^{\circ}$26$'$31\farcs 3 & 0.13 & BLG657.24 & 3809   & 18.741 & 19.424 & 0.683 & MS? & 0.1237444(2) \\
\object{V5627 Sgr} & --        & 18$^{\rm{h}}$01$^{\rm{m}}$56\fs 26 & -27$^{\circ}$22$'$56\farcs 1 & 0.39 & BLG233.8 & 22490  & 15.994 & 17.265 & 1.271 & MS & 0.117161(2) \\
\object{V825 Sco} & Sco 1964  & 17$^{\rm{h}}$49$^{\rm{m}}$53\fs 78 & -33$^{\circ}$32$'$13\farcs 0 & 1.81 & BLG502.14 & 149263 & 18.232 & 19.537 & 1.304 & MS & 0.19165876(3)\\
\hline
\end{tabular} 
\begin{flushleft}
Note. For each object, we give designation, equatorial coordinates (for the epoch J2000.0), distance from the SIMBAD position, the OGLE identifier, mean $I$- and $V$-band brightness, $V-I$ color, the type of the secondary (MS - main sequence, RG - red giant), and orbital period.
\end{flushleft}
\label{tab:ecl}
\end{table*}

\begin{table*}
\caption{Possible post-novae: old novae counterparts. For each object, we give designation, equatorial coordinates (for the epoch J2000.0), distance from the SIMBAD position, the OGLE identifier, mean $I$- and $V$-band brightness, $V-I$ color, and the type of the secondary (MS - main sequence, RG - red giant).}
\begin{tabular}{lllllllllll}
\hline
GCVS ID & Nova & RA$_{\rm J2000.0}$                      & DEC$_{\rm J2000.0}$            & d [$''$] & Field & Star ID & $I$ & $V$ & $V-I$ & Donor type \\
\hline

\object{V553 Oph} & Oph 1940  & 17$^{\rm{h}}$42$^{\rm{m}}$53\fs 47 & -24$^{\circ}$51$'$25\farcs 5 & 0.75 & BLG633.16 & 78266  & 18.830 & 21.382 & 2.552 & MS \\
\object{V794 Oph} & Oph 1983  & 17$^{\rm{h}}$38$^{\rm{m}}$49\fs 25 & -22$^{\circ}$50$'$49\farcs 0 & 0.14 & BLG625.14 & 88999  & 15.697 & 17.823 & 2.126 & RG \\
\object{V972 Oph} & Oph 1957  & 17$^{\rm{h}}$34$^{\rm{m}}$44\fs 46 & -28$^{\circ}$10$'$35\farcs 8 & 0.37 & BLG653.30 & 69132  & 15.076 & 16.627 & 1.551 & MS \\
\object{V1012 Oph} & Oph 1961  & 17$^{\rm{h}}$41$^{\rm{m}}$34\fs 44 & -23$^{\circ}$23$'$32\farcs 2 & 1.18 & BLG625.02 & 71413  & 18.793 & 20.509 & 1.715 & MS \\
\object{V2110 Oph} & Oph 1950  & 17$^{\rm{h}}$43$^{\rm{m}}$33\fs 32 & -22$^{\circ}$45$'$36\farcs 9 & 0.11 & BLG631.07 & 39420  & 13.880 & 17.673 & 3.792 & RG \\
\object{V2575 Oph} & Oph 2006  & 17$^{\rm{h}}$33$^{\rm{m}}$13\fs 06 & -24$^{\circ}$21$'$07\farcs 1 & 0.07 & BLG715.32 & 1542   & 19.340 & 21.540 & 2.199 & MS \\
\object{V2671 Oph} & Oph 2008b & 17$^{\rm{h}}$33$^{\rm{m}}$29\fs 70 & -27$^{\circ}$01$'$16\farcs 3 & 0.38 & BLG611.24 & 51799  & 21.336 & --     & --    & -- \\
\object{AT Sgr}    & --        & 18$^{\rm{h}}$03$^{\rm{m}}$30\fs 84 & -26$^{\circ}$28$'$29\farcs 2 & 0.74 & BLG646.03 & 92517  & 19.414 & --     & --    & -- \\
\object{FL Sgr}    & Sgr 1924  & 18$^{\rm{h}}$00$^{\rm{m}}$30\fs 28 & -34$^{\circ}$36$'$12\farcs 0 & 1.53 & BLG517.25 & 33354  & 18.301 & 19.456 & 1.156 & MS \\
\object{FM Sgr}    & Sgr 1926  & 18$^{\rm{h}}$17$^{\rm{m}}$18\fs 21 & -23$^{\circ}$38$'$28\farcs 2 & 0.70 & BLG527.07 & 1123   & 16.486 & 17.792 & 1.306 & MS \\
\object{GR Sgr}    & --        & 18$^{\rm{h}}$22$^{\rm{m}}$58\fs 50 & -25$^{\circ}$34$'$47\farcs 3 & 0.03 & BLG263.8  & 48728  & 15.768 & 16.298 & 0.531 & MS? \\
\object{HS Sgr}    & Sgr 1900  & 18$^{\rm{h}}$28$^{\rm{m}}$03\fs 49 & -21$^{\circ}$34$'$24\farcs 3 & 0.81 & BLG304.4  & 40060  & 18.803 & 19.506 & 0.702 & MS? \\
\object{KY Sgr}    & Sgr 1926a & 18$^{\rm{h}}$01$^{\rm{m}}$21\fs 02 & -26$^{\circ}$24$'$40\farcs 0 & 0.03 & BLG646.14 & 67578  & 14.719 & 18.870 & 4.151 & RG \\
\object{V441 Sgr} & Sgr 1930  & 18$^{\rm{h}}$22$^{\rm{m}}$08\fs 10 & -25$^{\circ}$28$'$54\farcs 2 & 0.57 & BLG529.17 & 5930   & 18.714 & 19.935 & 1.221 & MS \\
\object{V732 Sgr} & Sgr 1936  & 17$^{\rm{h}}$56$^{\rm{m}}$07\fs 51 & -27$^{\circ}$22$'$17\farcs 1 & 0.43 & BLG645.06 & 10991  & 15.376 & 16.902 & 1.527 & MS \\
\object{V737 Sgr} & Sgr 1933  & 18$^{\rm{h}}$07$^{\rm{m}}$08\fs 71 & -28$^{\circ}$44$'$52\farcs 3 & 0.66 & BLG520.31 & 115201 & 16.589 & 18.195 & 1.605 & RG? \\
\object{V787 Sgr} & Sgr 1937  & 18$^{\rm{h}}$00$^{\rm{m}}$02\fs 18 & -30$^{\circ}$30$'$29\farcs 9 & 0.07 & BLG183.6  & 192974 & 17.779 & 19.122 & 1.343 & MS \\
\object{V927 Sgr} & Sgr 1944  & 18$^{\rm{h}}$07$^{\rm{m}}$43\fs 16 & -33$^{\circ}$21$'$17\farcs 6 & 0.18 & BLG588.06 & 47399  & 17.901 & 19.391 & 1.490 & RG? \\
\object{V928 Sgr} & Sgr 1947a & 18$^{\rm{h}}$19$^{\rm{m}}$00\fs 34 & -28$^{\circ}$06$'$00\farcs 8 & 0.87 & BLG531.13 & 48681  & 18.673 & 19.937 & 1.264 & MS \\
\object{V990 Sgr} & Sgr 1936d & 17$^{\rm{h}}$57$^{\rm{m}}$18\fs 24 & -28$^{\circ}$19$'$07\farcs 9 & 0.15 & BLG214.7  & 223677 & 15.083 & 18.126 & 3.043 & RG \\
\object{V1012 Sgr} & Sgr 1914c & 18$^{\rm{h}}$06$^{\rm{m}}$14\fs 10 & -31$^{\circ}$44$'$28\farcs 1 & 0.06 & BLG573.16 & 29456  & 15.676 & 17.398 & 1.721 & RG \\
\object{V1014 Sgr} & Sgr 1901  & 18$^{\rm{h}}$06$^{\rm{m}}$45\fs 90 & -27$^{\circ}$26$'$16\farcs 3 & 0.24 & BLG226.5  & 177378 & 18.425 & 19.598 & 1.173 & MS? \\
\object{V1015 Sgr} & Sgr 1905  & 18$^{\rm{h}}$09$^{\rm{m}}$01\fs 95 & -32$^{\circ}$28$'$33\farcs 0 & 0.97 & BLG588.29 & 79153  & 19.799 & 21.509 & 1.710 & MS \\
\object{V1148 Sgr} & Sgr 1943  & 18$^{\rm{h}}$09$^{\rm{m}}$05\fs 85 & -25$^{\circ}$59$'$08\farcs 0 & 0.07 & BLG580.03 & 65227  & 14.015 & 14.865 & 0.849 & MS \\
\object{V1149 Sgr} & Sgr 1945a & 18$^{\rm{h}}$18$^{\rm{m}}$30\fs 51 & -28$^{\circ}$17$'$16\farcs 8 & 0.52 & BLG531.13 & 84838  & 17.897 & 19.025 & 1.128 & MS \\
\object{V1150 Sgr} & Sgr 1948  & 18$^{\rm{h}}$18$^{\rm{m}}$55\fs 40 & -24$^{\circ}$05$'$33\farcs 0 & 1.28 & BLG528.21 & 132239 & 19.772 & 21.518 & 1.746 & MS? \\
\object{V1172 Sgr} & Sgr 1951a & 17$^{\rm{h}}$50$^{\rm{m}}$23\fs 66 & -20$^{\circ}$40$'$29\farcs 8 & 0.12 & BLG636.12 & 40656  & 14.859 & 18.042 & 3.183 & RG \\
\object{V1274 Sgr} & Sgr 1954a & 17$^{\rm{h}}$48$^{\rm{m}}$55\fs 52 & -17$^{\circ}$52$'$02\farcs 5 & 1.88 & BLG353.8  & 62657  & 19.794 & 21.244 & 1.450 & MS \\
\object{V1275 Sgr} & Sgr 1954b & 17$^{\rm{h}}$59$^{\rm{m}}$06\fs 35 & -36$^{\circ}$18$'$40\farcs 8 & 0.03 & BLG536.19 & 37780  & 15.441 & --     & --    & -- \\
\object{V1431 Sgr} & Sgr 1945b & 18$^{\rm{h}}$03$^{\rm{m}}$29\fs 55 & -29$^{\circ}$59$'$54\farcs 8 & 0.76 & BLG513.11 & 132930 & 14.144 & 15.988 & 1.844 & RG \\
\object{V1583 Sgr} & Sgr 1928  & 18$^{\rm{h}}$15$^{\rm{m}}$26\fs 52 & -23$^{\circ}$23$'$17\farcs 6 & 0.71 & BLG279.5  & 19002  & 19.465 & -- & -- & -- \\
\object{V1944 Sgr} & Sgr 1960  & 18$^{\rm{h}}$00$^{\rm{m}}$36\fs 80 & -27$^{\circ}$17$'$14\farcs 4 & 0.72 & BLG511.24 & 135231 & 18.518 & 20.373 & 1.855 & MS \\
\object{V2415 Sgr} & Sgr 1951b & 17$^{\rm{h}}$53$^{\rm{m}}$11\fs 64 & -29$^{\circ}$34$'$25\farcs 5 & 0.68 & BLG501.19 & 161372 & 17.968 & --     & --    & -- \\
\object{V3889 Sgr} & Sco 1975  & 17$^{\rm{h}}$58$^{\rm{m}}$21\fs 26 & -28$^{\circ}$21$'$52\farcs 8 & 0.76 & BLG504.03 & 95425  & 19.180 & 20.860 & 1.679 & MS? \\
\object{V3890 Sgr} & Sgr 1962  & 18$^{\rm{h}}$30$^{\rm{m}}$43\fs 28 & -24$^{\circ}$01$'$08\farcs 9 & 0.07 & BLG278.6  & 85802  & 12.530 & 15.944 & 3.413 & RG \\
\object{V3964 Sgr} & Sgr 1975b & 17$^{\rm{h}}$49$^{\rm{m}}$42\fs 63 & -17$^{\circ}$23$'$36\farcs 3 & 0.11 & BLG353.4  & 74779  & 17.153 & 18.739 & 1.587 & MS \\
\object{V4027 Sgr} & Sgr 1968  & 18$^{\rm{h}}$02$^{\rm{m}}$29\fs 24 & -28$^{\circ}$45$'$19\farcs 2 & 0.79 & BLG512.13 & 87364  & 18.477 & 19.939 & 1.462 & MS \\
\object{V4049 Sgr} & Sgr 1978  & 18$^{\rm{h}}$20$^{\rm{m}}$38\fs 03 & -27$^{\circ}$56$'$26\farcs 0 & 0.41 & BLG531.19 & 83212  & 18.986 & 20.227 & 1.241 & MS \\
\object{V4065 Sgr} & Sgr 1980  & 18$^{\rm{h}}$19$^{\rm{m}}$38\fs 20 & -24$^{\circ}$43$'$56\farcs 0 & 0.13 & BLG528.03 & 103076 & 12.496 & 14.224 & 1.728 & RG? \\
\object{V4092 Sgr} & Sgr 1984  & 17$^{\rm{h}}$53$^{\rm{m}}$41\fs 99 & -29$^{\circ}$02$'$08\farcs 3 & 0.11 & BLG500.02 & 35755  & 18.923 & 20.970 & 2.047 & MS \\
\object{V4121 Sgr} & Sgr 1983  & 18$^{\rm{h}}$07$^{\rm{m}}$54\fs 79 & -28$^{\circ}$49$'$27\farcs 8 & 1.07 & BLG520.30 & 60926  & 18.881 & 19.756 & 0.875 & MS? \\
\object{V4135 Sgr} & Sgr 1987  & 17$^{\rm{h}}$59$^{\rm{m}}$45\fs 14 & -32$^{\circ}$16$'$20\farcs 6 & 0.75 & BLG507.01 & 37851  & 18.935 & 21.069 & 2.134 & MS \\
\object{V4157 Sgr} & Sgr 1992a & 18$^{\rm{h}}$09$^{\rm{m}}$35\fs 03 & -25$^{\circ}$51$'$58\farcs 8 & 0.63 & BLG580.02 & 128837 & 19.341 & 20.820 & 1.479 & MS \\
\object{V4160 Sgr} & Sgr 1991  & 18$^{\rm{h}}$14$^{\rm{m}}$13\fs 69 & -32$^{\circ}$12$'$28\farcs 4 & 1.80 & BLG546.22 & 13907  & 19.177 & 20.385 & 1.208 & MS \\
\object{V4171 Sgr} & Sgr 1992c & 18$^{\rm{h}}$23$^{\rm{m}}$41\fs 39 & -22$^{\circ}$59$'$29\farcs 0 & 0.69 & BLG289.2  & 92352  & 19.434 & 20.938 & 1.505 & MS \\
\object{V4327 Sgr} & Sgr 1993  & 18$^{\rm{h}}$12$^{\rm{m}}$49\fs 76 & -29$^{\circ}$29$'$05\farcs 4 & 0.98 & BLG525.31 & 47019  & 19.371 & 21.069 & 1.698 & MS? \\
\object{V4338 Sgr} & Sgr 1990  & 17$^{\rm{h}}$59$^{\rm{m}}$18\fs 17 & -29$^{\circ}$09$'$53\farcs 0 & 0.11 & BLG505.19 & 10131  & 18.527 & --     & --    & -- \\
\object{V4643 Sgr} & Sgr 2001  & 17$^{\rm{h}}$54$^{\rm{m}}$40\fs 42 & -26$^{\circ}$14$'$15\farcs 5 & 0.07 & BLG647.08 & 11678  & 14.567 & 16.561 & 1.994 & MS \\
\object{V4744 Sgr} & Sgr 2002d & 17$^{\rm{h}}$47$^{\rm{m}}$21\fs 71 & -23$^{\circ}$28$'$23\farcs 0 & 0.46 & BLG632.09 & 158291 & 20.327 & 22.027 & 1.701 & MS \\
\object{V5114 Sgr} & Sgr 2004  & 18$^{\rm{h}}$19$^{\rm{m}}$32\fs 18 & -28$^{\circ}$36$'$35\farcs 4 & 1.46 & BLG531.04 & 3875   & 19.374 & 20.041 & 0.667 & MS? \\
\object{V5114 Sgr} & Sgr 2004  & 18$^{\rm{h}}$19$^{\rm{m}}$32\fs 18 & -28$^{\circ}$36$'$35\farcs 3 & 1.46 & BLG532.29 & 120953 & 19.343 & 19.997 & 0.653 & MS? \\
\object{V5115 Sgr} & Sgr 2005a & 18$^{\rm{h}}$16$^{\rm{m}}$59\fs 01 & -25$^{\circ}$56$'$38\farcs 6 & 0.43 & BLG529.07 & 159171 & 19.889 & --     & --    & -- \\
\object{V5117 Sgr} & Sgr 2006  & 17$^{\rm{h}}$58$^{\rm{m}}$52\fs 59 & -36$^{\circ}$47$'$34\farcs 6 & 0.59 & BLG536.10 & 60402  & 19.520 & 20.238 & 0.718 & MS? \\
\object{V5557 Sgr} & Sgr 1893  & 18$^{\rm{h}}$01$^{\rm{m}}$43\fs 14 & -35$^{\circ}$39$'$28\farcs 9 & 1.08 & BLG539.31 & 55072  & 18.488 & 19.999 & 1.511 & MS? \\
\object{V5581 Sgr} & Sgr 2009a & 17$^{\rm{h}}$44$^{\rm{m}}$08\fs 46 & -26$^{\circ}$05$'$48\farcs 0 & 0.81 & BLG652.17 & 54     & 14.209 & 19.757 & 5.548 & RG \\
\hline
\end{tabular} 
\label{tab:postnowe1}
\end{table*}

\addtocounter{table}{-1}

\begin{table*}
\caption{(Continued)}
\begin{tabular}{lllllllllll}
\hline
GCVS ID & Nova & RA$_{\rm J2000.0}$                      & DEC$_{\rm J2000.0}$            & d [$''$] & Field & Star ID & $I$ & $V$ & $V-I$ & Donor type \\
\hline
\object{KP Sco}    & Sco 1928  & 17$^{\rm{h}}$44$^{\rm{m}}$16\fs 55 & -35$^{\circ}$43$'$22\farcs 8 & 1.22 & BLG604.06 & 58473  & 20.510 & 21.866 & 1.355 & MS? \\
\object{V382 Sco} & Sco 1901  & 17$^{\rm{h}}$51$^{\rm{m}}$56\fs 17 & -35$^{\circ}$25$'$05\farcs 4 & 0.52 & BLG599.28 & 141030 & 20.069 & --     & --    & -- \\
\object{V696 Sco} & Sco 1944  & 17$^{\rm{h}}$53$^{\rm{m}}$11\fs 57 & -35$^{\circ}$50$'$14\farcs 4 & 0.09 & BLG599.19 & 34922  & 15.269 & 16.162 & 0.893 & MS \\
\object{V697 Sco} & Sco 1941  & 17$^{\rm{h}}$51$^{\rm{m}}$21\fs 79 & -37$^{\circ}$24$'$55\farcs 0 & 0.48 & BLG600.12 & 82272  & 20.938 & --     & --    & -- \\
\object{V707 Sco} & Sco 1922  & 17$^{\rm{h}}$48$^{\rm{m}}$26\fs 45 & -36$^{\circ}$37$'$54\farcs 7 & 0.93 & BLG605.08 & 121534 & 19.879 & 21.190 & 1.311 & MS? \\
\object{V711 Sco} & Sco 1906  & 17$^{\rm{h}}$54$^{\rm{m}}$06\fs 16 & -34$^{\circ}$21$'$15\farcs 6 & 0.08 & BLG503.26 & 8401   & 15.815 & 17.360 & 1.545 & RG? \\
\object{V719 Sco} & Sco 1950a & 17$^{\rm{h}}$45$^{\rm{m}}$43\fs 87 & -34$^{\circ}$00$'$55\farcs 0 & 0.04 & BLG603.21 & 102038 & 19.492 & --     & --    & -- \\
\object{V720 Sco} & Sco 1950b & 17$^{\rm{h}}$51$^{\rm{m}}$58\fs 13 & -35$^{\circ}$23$'$25\farcs 9 & 1.40 & BLG599.28 & 143222 & 19.917 & 21.474 & 1.557 & MS \\
\object{V722 Sco} & Sco 1952a & 17$^{\rm{h}}$48$^{\rm{m}}$36\fs 91 & -34$^{\circ}$57$'$55\farcs 2 & 2.15 & BLG503.16 & 52481  & 19.577 & 21.822 & 2.245 & MS \\
\object{V723 Sco} & Sco 1952b & 17$^{\rm{h}}$50$^{\rm{m}}$05\fs 32 & -35$^{\circ}$23$'$57\farcs 0 & 1.07 & BLG599.31 & 58786  & 17.061 & 18.978 & 1.917 & RG \\
\object{V733 Sco} & Sco 1937a & 17$^{\rm{h}}$39$^{\rm{m}}$43\fs 01 & -35$^{\circ}$52$'$38\farcs 5 & 1.63 & BLG610.21 & 40104  & 16.530 & 18.928 & 2.398 & RG \\
\object{V744 Sco} & Sco 1935  & 17$^{\rm{h}}$53$^{\rm{m}}$18\fs 13 & -31$^{\circ}$13$'$34\farcs 2 & 1.14 & BLG534.10 & 66288  & 20.700 & --     & --    & -- \\
\object{V902 Sco} & Sco 1949  & 17$^{\rm{h}}$26$^{\rm{m}}$08\fs 48 & -39$^{\circ}$03$'$59\farcs 2 & 1.26 & BLG201.5  & 66510  & 20.720 & --     & --    & -- \\
\object{V960 Sco} & Sco 1985  & 17$^{\rm{h}}$56$^{\rm{m}}$34\fs 14 & -31$^{\circ}$49$'$36\farcs 5 & 0.22 & BLG507.13 & 160149 & 19.625 & --     & --    & -- \\
\object{V977 Sco} & Sco 1989b & 17$^{\rm{h}}$51$^{\rm{m}}$50\fs 37 & -32$^{\circ}$31$'$57\farcs 5 & 0.11 & BLG535.12 & 45711  & 16.060 & 18.427 & 2.367 & RG \\
\object{V1142 Sco} & Sco 1998  & 17$^{\rm{h}}$55$^{\rm{m}}$25\fs 00 & -31$^{\circ}$01$'$41\farcs 1 & 0.44 & BLG506.07 & 54628  & 18.544 & --     & --    & -- \\
\object{V1187 Sco} & Sco 2004b & 17$^{\rm{h}}$29$^{\rm{m}}$18\fs 83 & -31$^{\circ}$46$'$01\farcs 5 & 0.23 & BLG668.13 & 11466  & 16.348 & 18.823 & 2.475 & MS? \\
\object{V1324 Sco} & Sco 2012  & 17$^{\rm{h}}$50$^{\rm{m}}$53\fs 92 & -32$^{\circ}$37$'$21\farcs 0 & 0.57 & BLG155.8  & 160743 & 19.240 & --     & --    & -- \\
\hline
\end{tabular}
\end{table*}

\begin{table}
\centering
\caption{Post-novae showing semi-regular variability.}
\begin{tabular}{lrr}
\hline
ID & Period [d] & Donor type$\dagger$ \\
\hline
V972 Oph  & 233          & MS \\
V2110 Oph & 192.8        & RG \\
GR Sgr    & 225          & MS? \\
V3890 Sgr & 104.5        & RG \\
V4643 Sgr & 32           & MS \\
V5581 Sgr & 62.3         & RG \\
V745 Sco  & 136.5; 77.4$\ddagger$  & RG \\
V1187 Sco & 354          & MS? \\
\hline
\end{tabular} \\
\begin{flushleft}
$\dagger$ Based on the location on the color-magnitude diagram: \\MS - main sequence, RG - red giant. \\
{\bf $\ddagger$ These two periods likely correspond to the fundamental-mode and 
first overtone radial pulsations of the red giant secondary (Mr\'oz et al. 2014). }
\end{flushleft}
\label{tab:SRVS}
\end{table}

\renewcommand{\thefigure}{A\arabic{figure}}
\setcounter{figure}{0}

\begin{appendix}

Figure \ref{fig:lcs1} presents the $I$-band light curves of all 39 CNe eruptions observed by the OGLE survey in the Galactic bulge. All objects are in lexicographical order.

\begin{figure*}
\centering
\begin{tabular}{cc}
\includegraphics[width=0.38\textwidth]{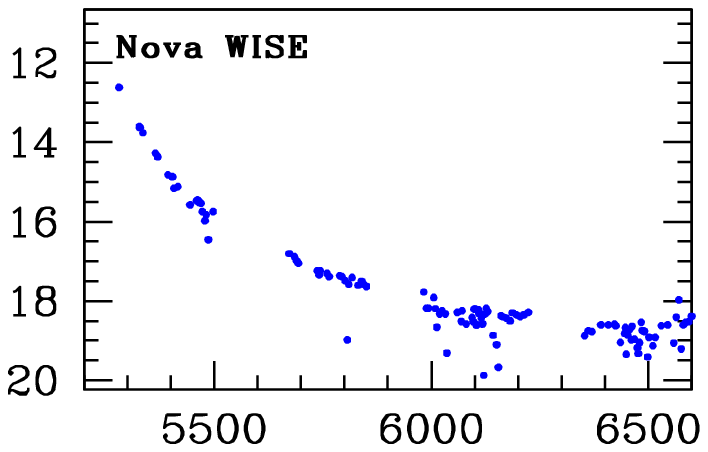} &
\includegraphics[width=0.38\textwidth]{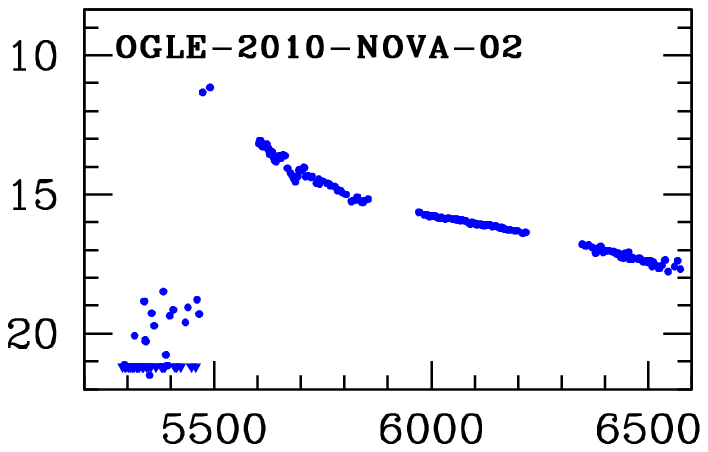} \\
\includegraphics[width=0.38\textwidth]{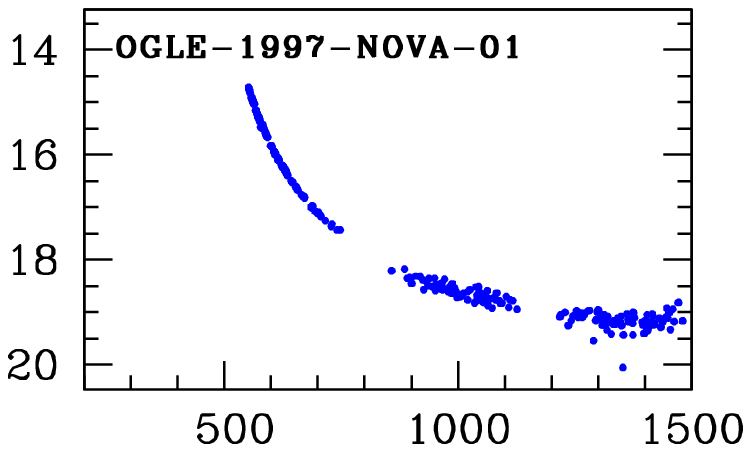} & 
\includegraphics[width=0.38\textwidth]{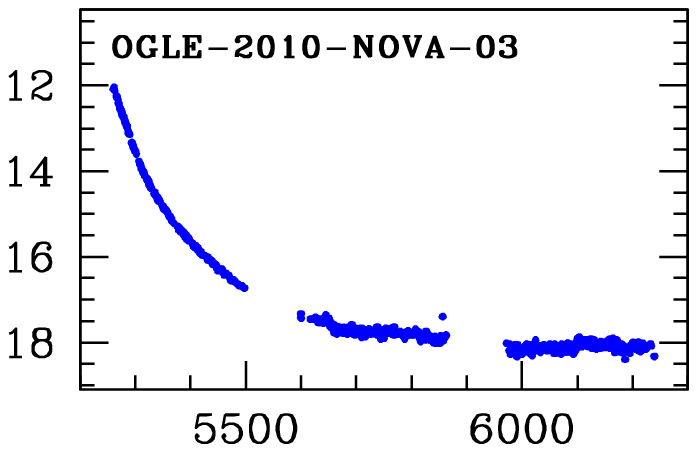} \\
\includegraphics[width=0.38\textwidth]{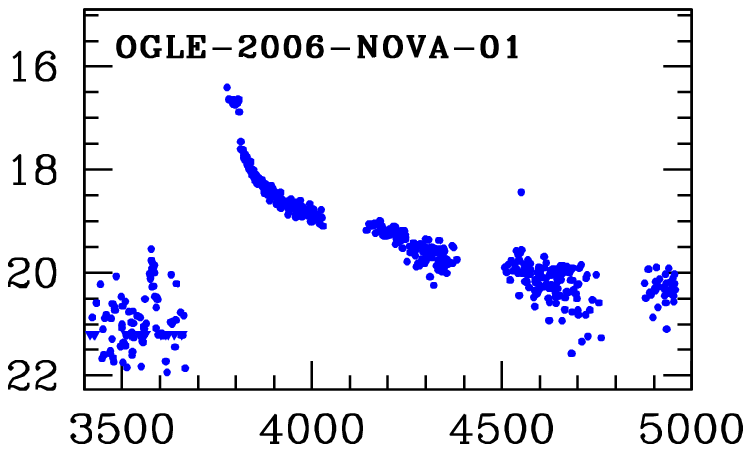} &
\includegraphics[width=0.38\textwidth]{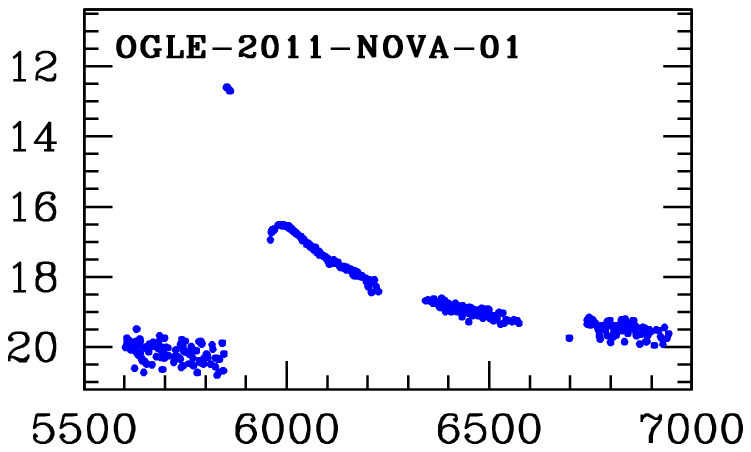} \\
\includegraphics[width=0.38\textwidth]{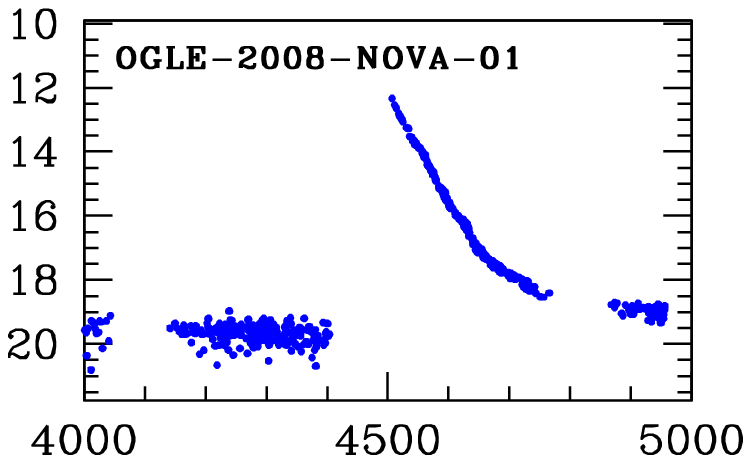} &
\includegraphics[width=0.38\textwidth]{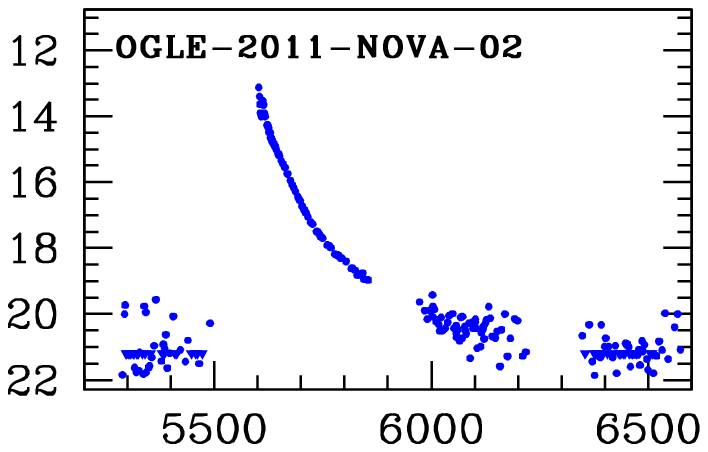} \\
\includegraphics[width=0.38\textwidth]{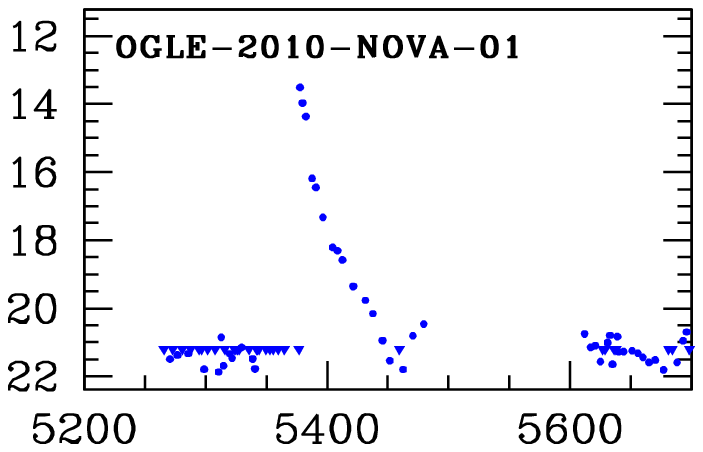}  &
\includegraphics[width=0.38\textwidth]{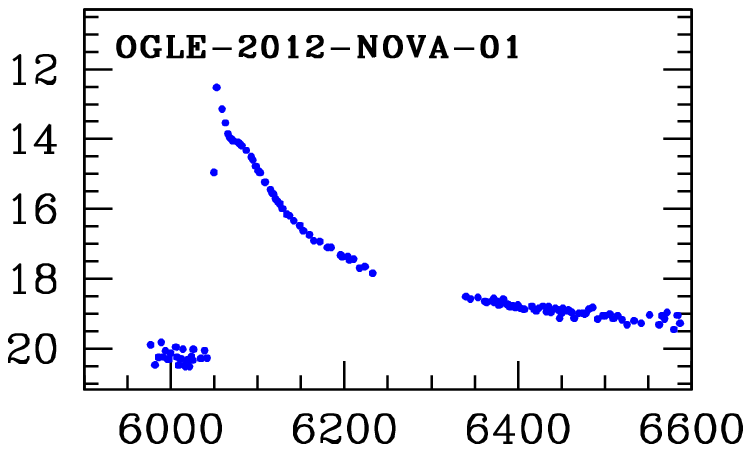} \\
\end{tabular}
\caption{Light curves of nova eruptions. Time (in the Heliocentric Julian Date minus 2450000) is on the $x$ axis, the $I$-band brightness is on the $y$ axis.}
\label{fig:lcs1}
\end{figure*}

\addtocounter{figure}{-1}

\begin{figure*}
\centering
\begin{tabular}{cc}
\includegraphics[width=0.38\textwidth]{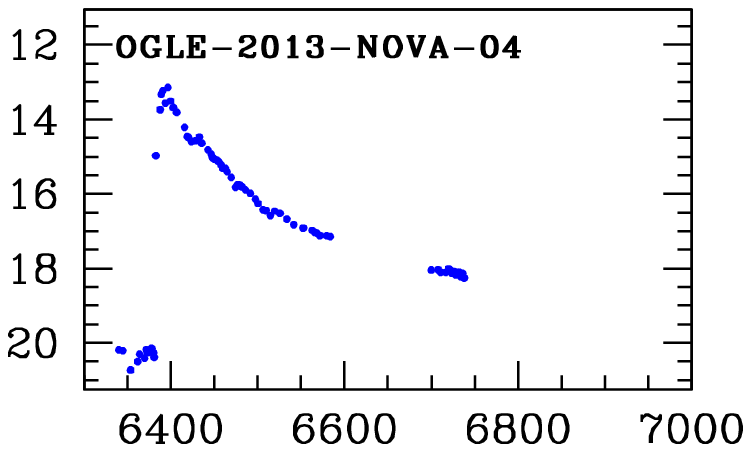} &
\includegraphics[width=0.38\textwidth]{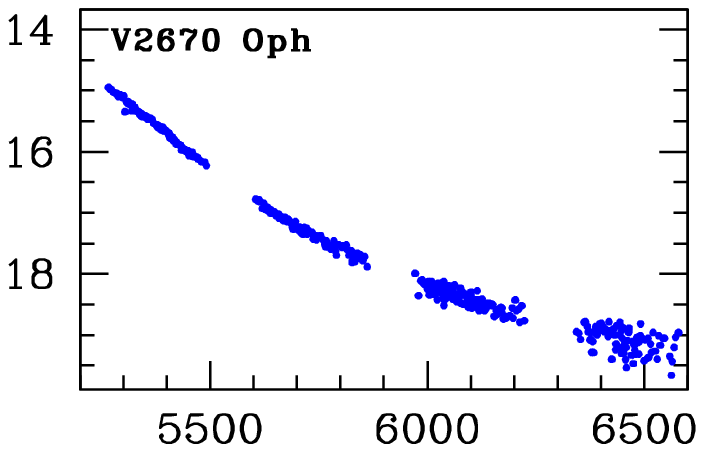} \\
\includegraphics[width=0.38\textwidth]{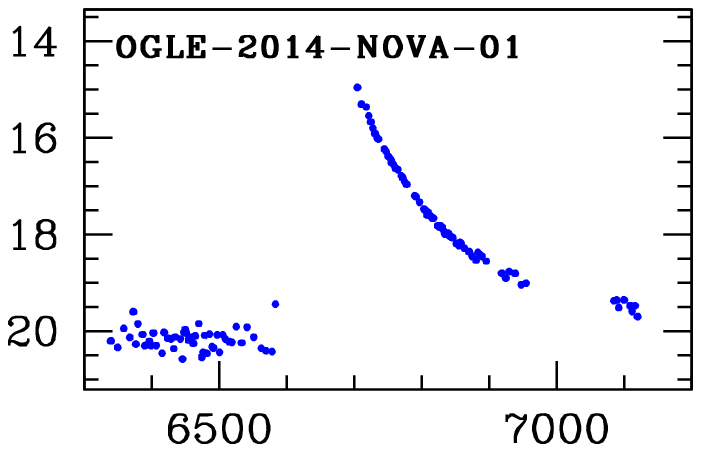} &
\includegraphics[width=0.38\textwidth]{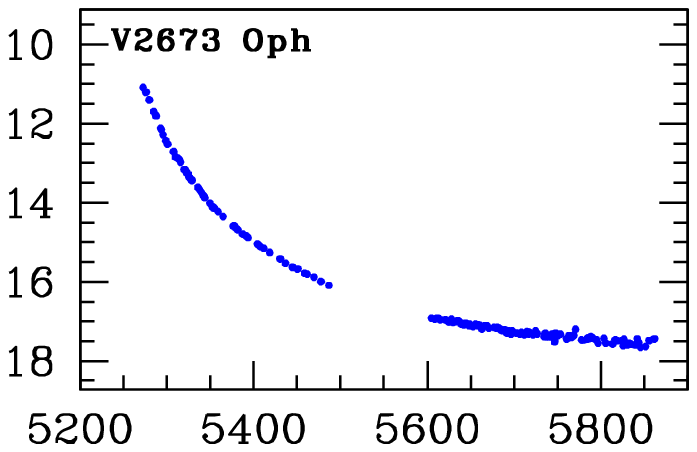} \\
\includegraphics[width=0.38\textwidth]{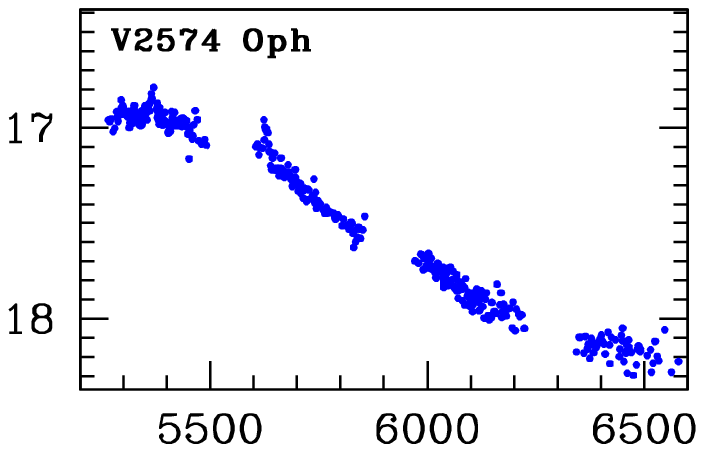} &
\includegraphics[width=0.38\textwidth]{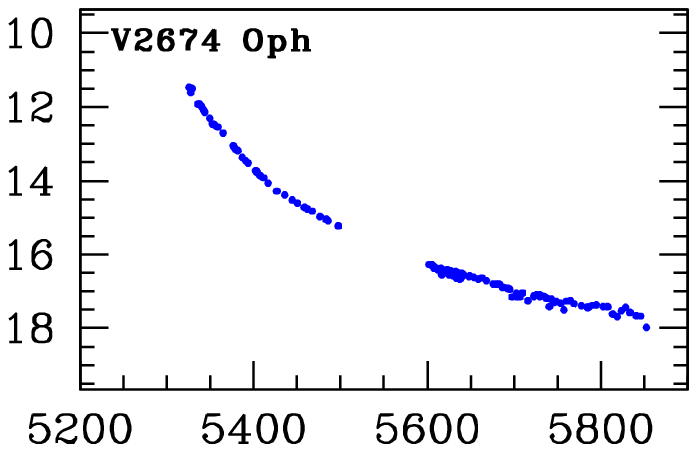} \\
\includegraphics[width=0.38\textwidth]{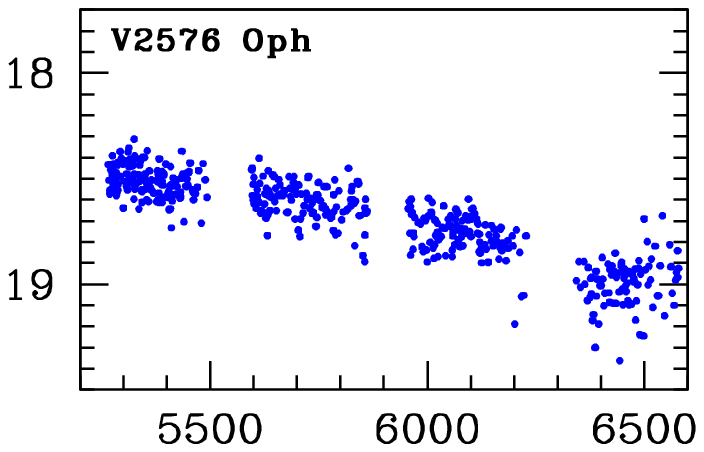} &
\includegraphics[width=0.38\textwidth]{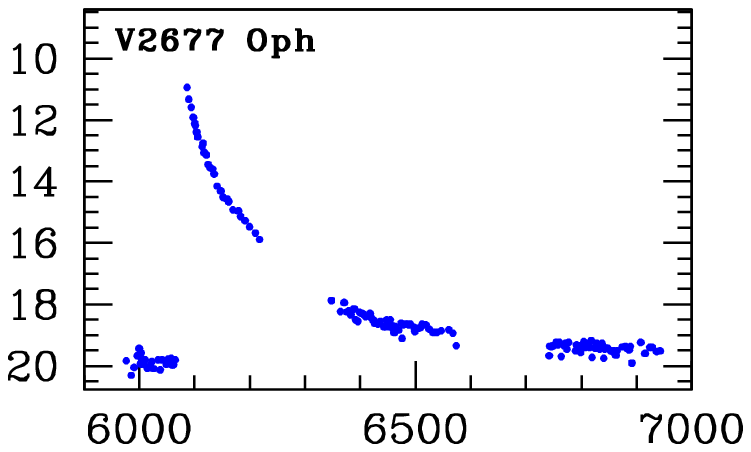} \\
\includegraphics[width=0.38\textwidth]{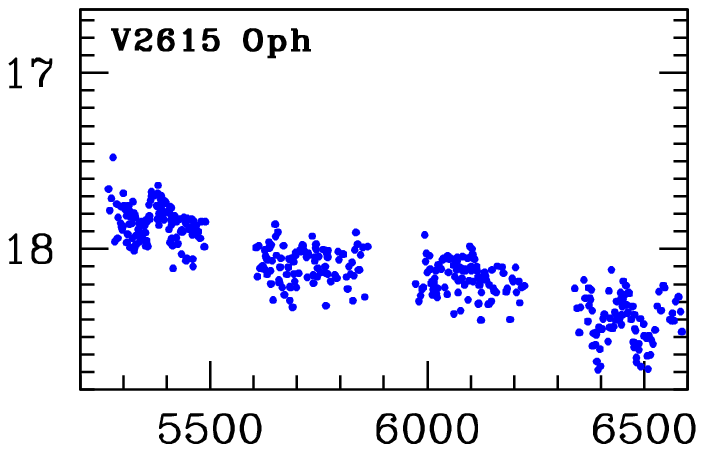} &
\includegraphics[width=0.38\textwidth]{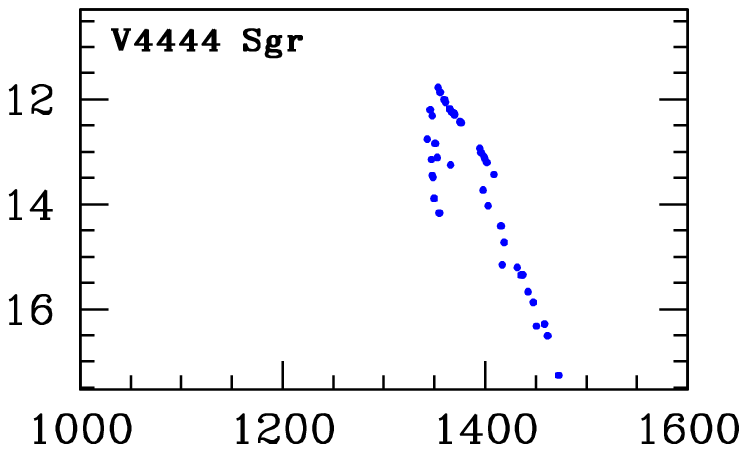} \\
\end{tabular}
\caption{(Continued)}
\end{figure*}

\addtocounter{figure}{-1}

\begin{figure*}
\centering
\begin{tabular}{cc}
\includegraphics[width=0.38\textwidth]{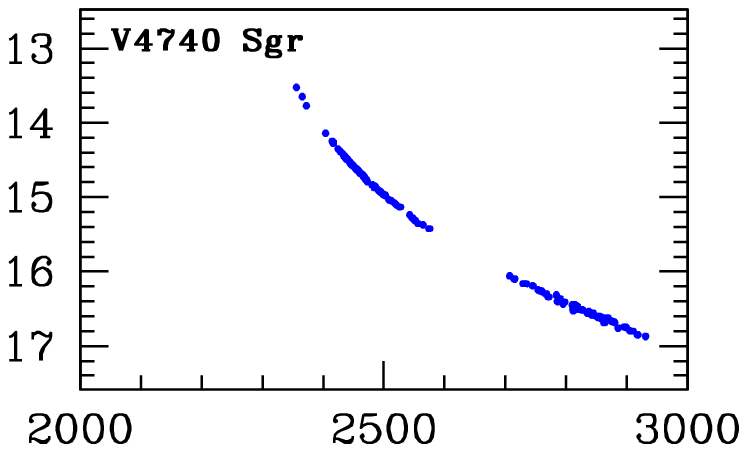} &
\includegraphics[width=0.38\textwidth]{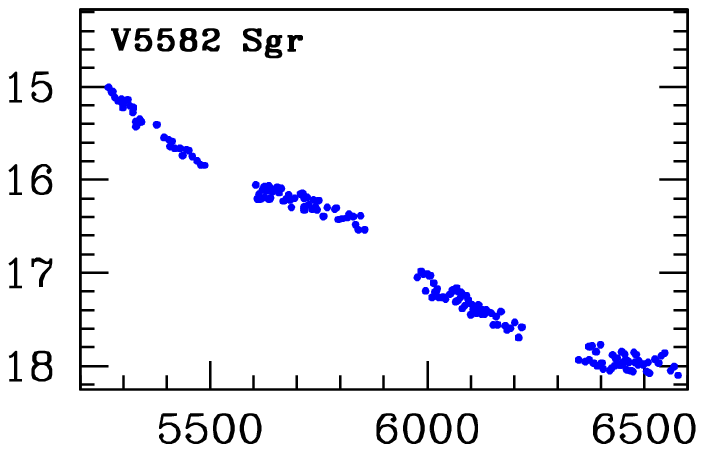} \\
\includegraphics[width=0.38\textwidth]{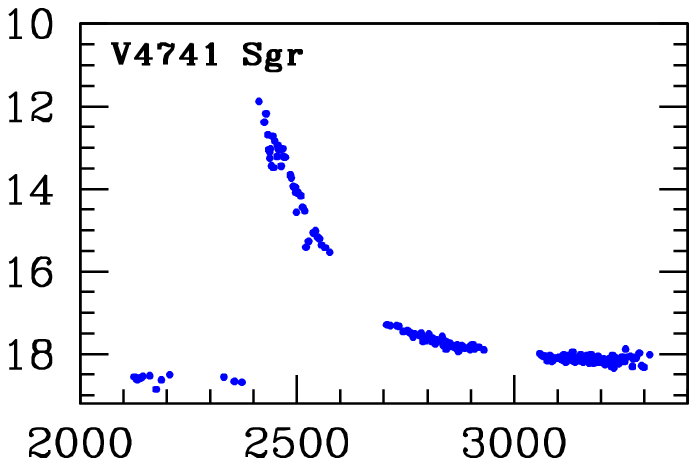} &
\includegraphics[width=0.38\textwidth]{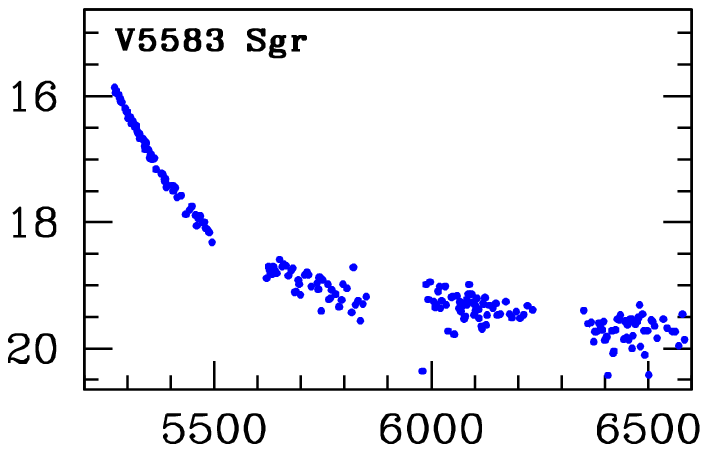} \\
\includegraphics[width=0.38\textwidth]{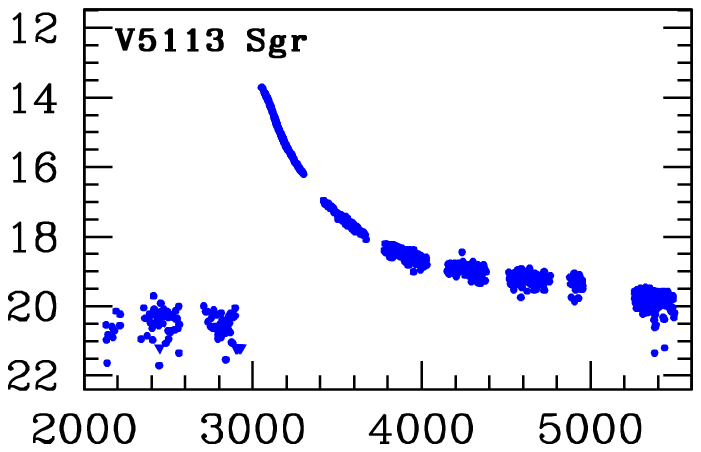} &
\includegraphics[width=0.38\textwidth]{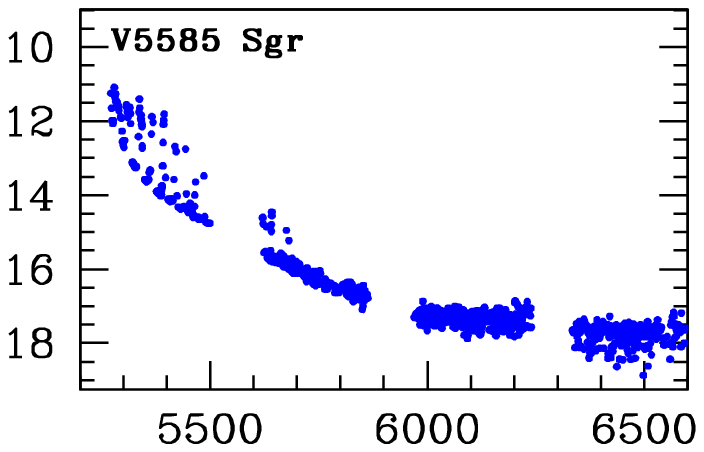} \\
\includegraphics[width=0.38\textwidth]{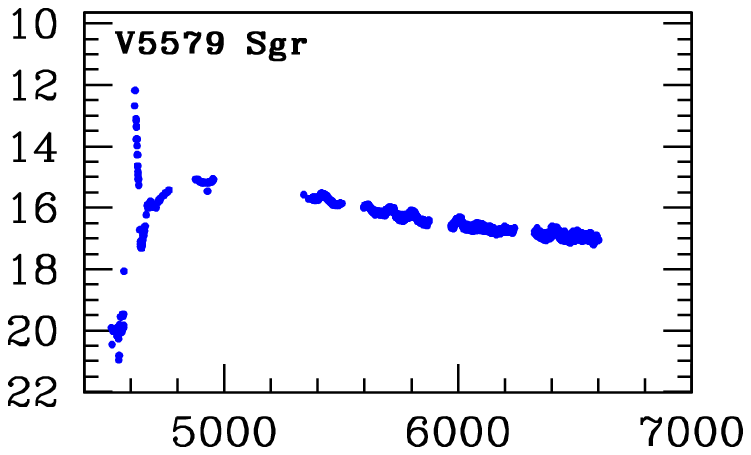} &
\includegraphics[width=0.38\textwidth]{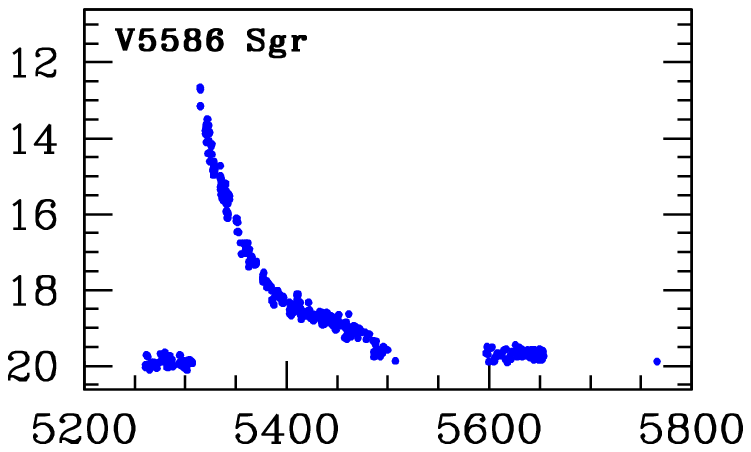} \\
\includegraphics[width=0.38\textwidth]{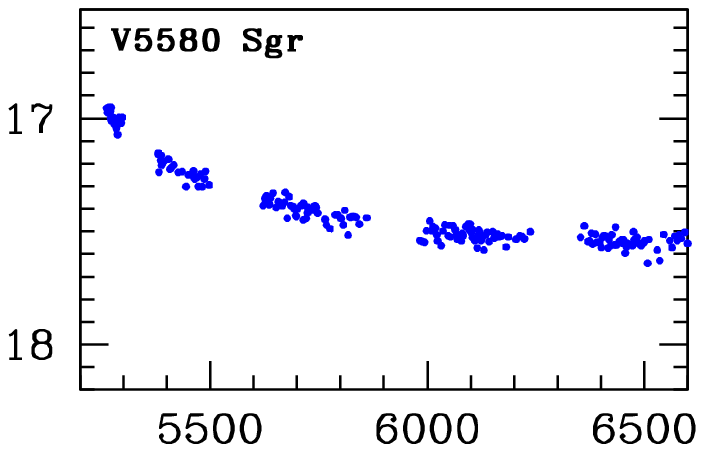} &
\includegraphics[width=0.38\textwidth]{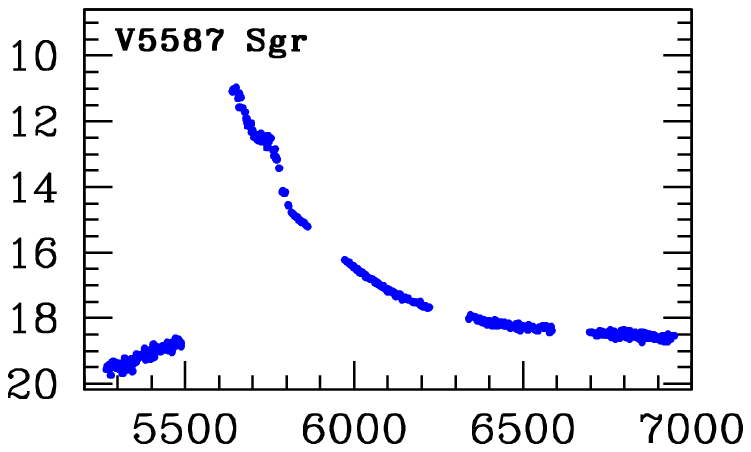} \\
\end{tabular}
\caption{(Continued)}
\end{figure*}

\addtocounter{figure}{-1}

\begin{figure*}
\centering
\begin{tabular}{cc}
\includegraphics[width=0.38\textwidth]{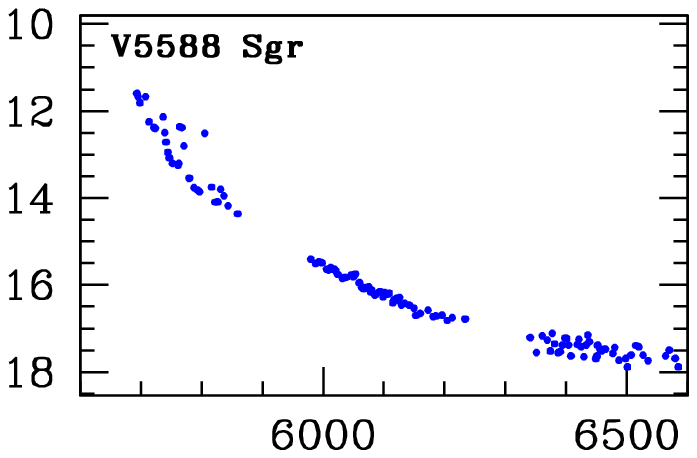} &
\includegraphics[width=0.38\textwidth]{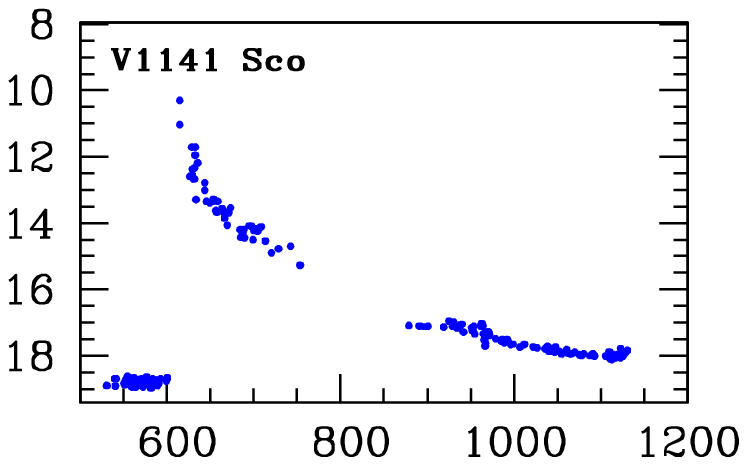} \\
\includegraphics[width=0.38\textwidth]{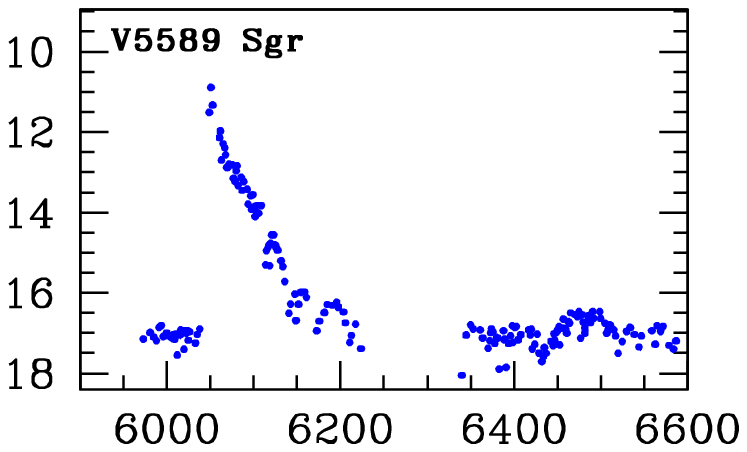} &
\includegraphics[width=0.38\textwidth]{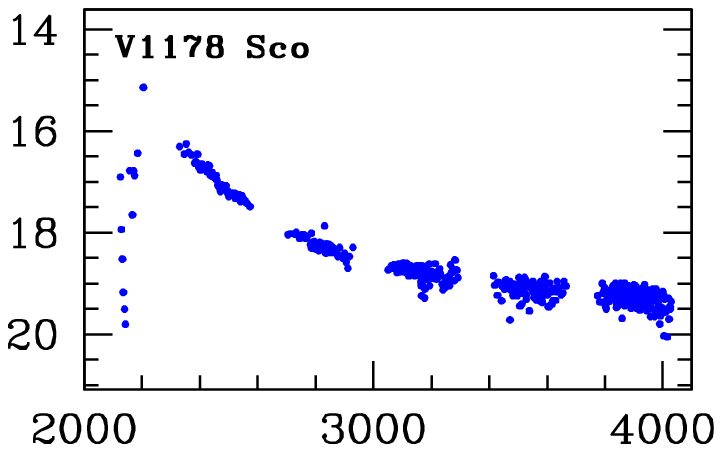} \\
\includegraphics[width=0.38\textwidth]{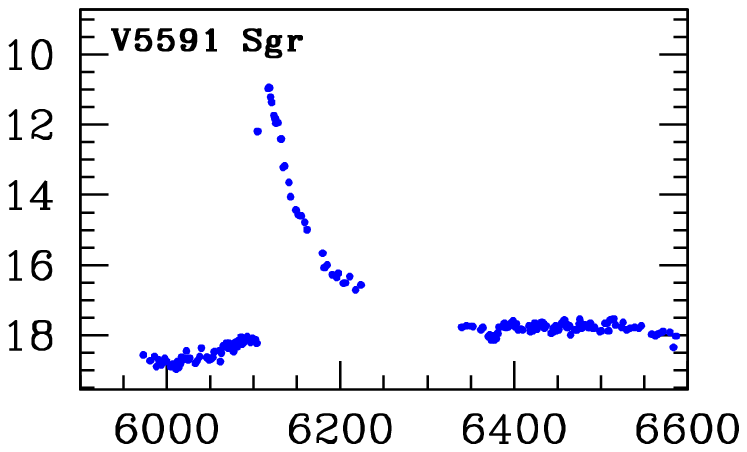} &
\includegraphics[width=0.38\textwidth]{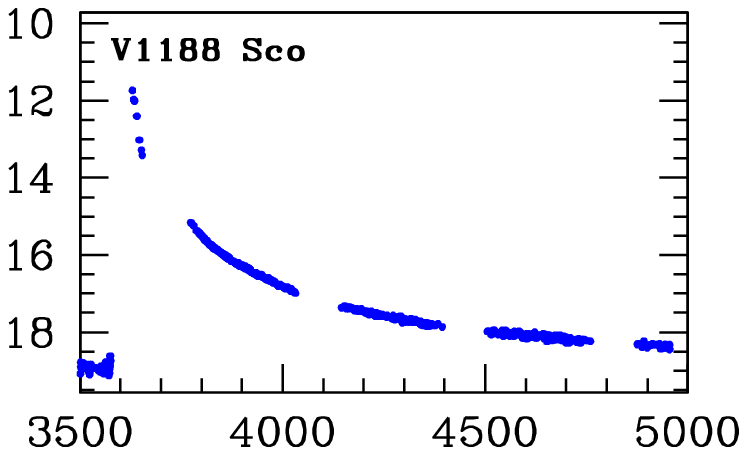} \\
\includegraphics[width=0.38\textwidth]{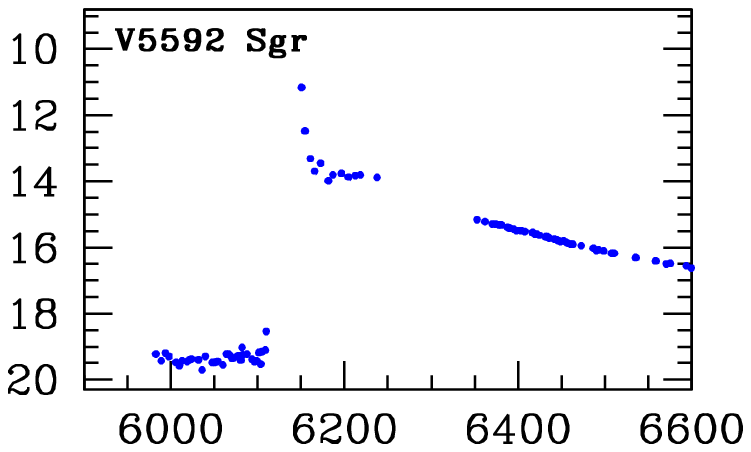} &
\includegraphics[width=0.38\textwidth]{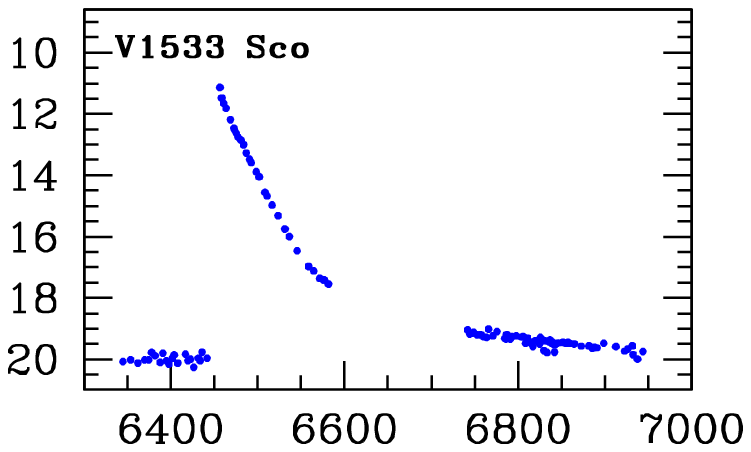} \\
\includegraphics[width=0.38\textwidth]{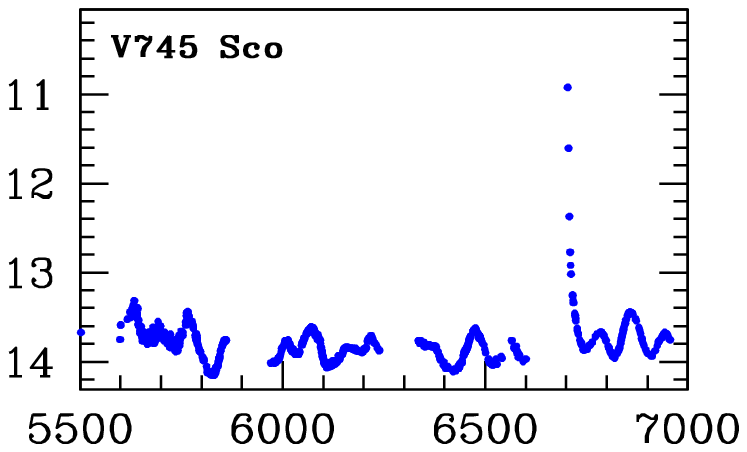}  &
\end{tabular}
\caption{(Continued)}
\end{figure*}

\end{appendix}

\end{document}